\documentclass[11pt]{article}

\usepackage{amsmath}
\usepackage{amssymb}
\usepackage{theorem}
\usepackage{citesort}
\usepackage{vmargin}

\setpapersize{USletter}
\setmarginsrb{1in}{1in}{1in}{1in}{0in}{0in}{\footheight}{0.5in}

\newcommand\abs[1]{\left|#1\right|}
\newcommand\Set[2]{\{#1 \, | \, #2 \}}
\newcommand\set[1]{\{ #1 \}}
\newcommand\R{\mathbb{R}}
\newcommand\Oh{\mathcal{O}}
\newcommand\oh{\mathrm{o}}
\newcommand\Expect{\mathbb{E}}
\newcommand\E\Expect
\newcommand\filledSquare{\rule{7pt}{7pt}}

\theorembodyfont{\slshape}
\newtheorem{theorem}{Theorem}
\newtheorem{lemma}[theorem]{Lemma}
\newtheorem{corollary}[theorem]{Corollary}
{\theorembodyfont{\rmfamily} \newtheorem{remark}{Remark}}
\newenvironment{proof}{\par\noindent{\bf Proof: }}{\nopagebreak\hfill\filledSquare\medskip}
\newenvironment{farproof}[1]{\par\noindent{\bf Proof of #1: }}{\nopagebreak\hfill\filledSquare\medskip}
{\theorembodyfont{\rmfamily} }

\newcommand{\suppress}[1]{}

\newcommand{\etal}[1]{{\it et al.}}

\newcommand{\ie}{{\it i.e.}}
\newcommand{\etc}{{\it etc.}}

\newcommand{\veca}{\mathbf{a}}
\newcommand{\vecb}{\mathbf{b}}

\newcommand{\vecx}{\mathbf{x}}
\newcommand{\vecv}{\mathbf{v}}

\newcommand{\vecp}{\mathbf{p}}

\newcommand{\vecf}{\mathbf{f}}
\newcommand{\vecw}{\mathbf{w}}

\newcommand{\MA}{\mathrm{\text{MA}}} 
\newcommand{\SR}{\mathrm{\text{SR}}} 

\newcommand{\CRP}{\mathrm{\text{CRP}}} 
\newcommand{\CRPside}{\mathrm{\text{CRP-S}}} 
\newcommand{\IA}{\mathrm{\text{IA}}} 

\newcommand{\U}{\mathcal{U}}
\newcommand{\D}{\mathcal{D}}
\newcommand{\eff}{\psi}
\newcommand{\inc}{\phi}

\newcommand{\Ret}{\mathcal{R}}
\newcommand{\LogRet}{\mathcal{L}}
\newcommand{\timeInt}{\mathcal{I}}

\newcommand{\calW}{\mathcal{W}}
\newcommand{\bbW}{\mathbb{W}}

\newcommand{\env}{\mathcal{E}}
\newcommand{\Vol}{\mathrm{Vol}}

\begin{document}

\title{Fast Universalization of Investment Strategies with Provably
Good Relative Returns}
\author{Karhan Akcoglu\thanks{
Department of Computer Science, Yale University,
New Haven, CT 06520, USA.
Email: {\tt karhan.akcoglu@yale.edu}.
Supported in part by NSF Grant CCR-9988376.}
\and Petros Drineas\thanks{
Department of Computer Science, Yale University,
New Haven, CT 06520, USA.
Email: {\tt petros.drineas@yale.edu}.
Supported in part by NSF Grant CCR-9896165.}
\and
Ming-Yang Kao\thanks{Department of Computer Science, Northwestern University, 
Evanston, IL 60201,  Email: {\tt kao@cs.northwestern.edu}.
Supported in part by NSF Grant CCR-9988376.}
}

\maketitle

\begin{abstract}
A \emph{universalization} of a parameterized investment strategy is an
online algorithm whose average daily performance approaches that of
the strategy operating with the optimal parameters determined offline in hindsight. We present a general framework for universalizing
investment strategies and discuss conditions under which investment
strategies are universalizable. We present examples of common
investment strategies that fit into our framework. The examples include
both trading strategies that decide positions in individual stocks,
and portfolio strategies that allocate wealth among multiple stocks. This work extends Cover's universal portfolio work. We
also discuss the runtime
efficiency of universalization algorithms. While a straightforward
implementation of our algorithms runs in time exponential in the number of
parameters, we show that the efficient universal portfolio computation technique of Kalai and Vempala involving the sampling of log-concave functions can be generalized to other classes of investment strategies.
\end{abstract}

\section{Introduction}
\label{section-intro}

An age-old question in finance deals with how to manage money on the
stock market to obtain an ``acceptable'' return on investment. An
\emph{investment strategy} is an online algorithm that attempts to
address this question by applying a given set of rules to determine
how to invest capital. Typically, an investment strategy is
parameterized by a vector $\vecw \in \R^* = \bigcup_{i=1}^\infty \R^i$ 
that dictates how the
strategy operates. The optimal parameters that maximize the strategy's
return are unknown when the algorithm is run and the
parameters are usually chosen quite arbitrarily. A
\emph{universalization} of an investment strategy is an online
algorithm based on the strategy whose average daily performance
approaches that of the strategy operating with the optimal parameters
determined offline in hindsight.

Consider the \emph{constantly rebalanced portfolio} ($\CRP$) investment strategy universalized by Cover \cite{Cover:1991:UP} and the subject of several extensions and generalizations \cite{Cover:1996:UPS,Ordentlich:1996:UPS,Helmbold:1998:LPS,Blum:1999:UPT,Kalai:2000:EAU}. The $\CRP$ strategy maintains a constant proportion of total wealth in each stock, where the proportions are dictated by the parameters given to the strategy. In a stock market with $m$ stocks, the parameter space for the $\CRP$ strategy is
$$
\calW_m = \Set{\vecw\in[0,1]^m}{\sum_{i=1}^m w_i = 1},
$$
the set of vectors in $\R^m$ whose components are between $0$ and $1$ and add up to $1$. Given a \emph{portfolio vector} $\vecw = (w_1, \ldots, w_m) \in\calW_m$, $w_i$ tells us the proportion of wealth to invest in stock $i$, for $1\le i\le m$. At the beginning of each day, the holdings are \emph{rebalanced}, \ie, money is taken out of some stocks and put into others, so that the desired proportions are maintained in each stock. As an example of the robustness of the $\CRP$ strategy, consider the following market with two stocks \cite{Ordentlich:1996:UPS,Helmbold:1998:LPS}. The price of one stock remains constant, while the other stock doubles and halves in price on alternate days. Investing in a single stock will at most double our money. With a $\CRP(\frac{1}{2}, \frac{1}{2})$ strategy, our wealth will increase exponentially, by a factor of $(\frac{1}{2}\cdot 1 + \frac{1}{2}\cdot 2) \times (\frac{1}{2}\cdot 1 + \frac{1}{2}\cdot \frac{1}{2}) = \frac{3}{2}\times\frac{3}{4} = \frac{9}{8}$ every two days.

Cover developed an investment strategy that effectively distributes wealth uniformly over all portfolio vectors $\vecw\in\calW_m$ on the first day and executes the $\CRP$ strategy with daily rebalancing according to each $\vecw$ on the (infinitesimally small) proportion of wealth initially allocated to each $\vecw$.  Cover showed that the \emph{average daily log-performance}\footnote{The average daily log-performance is the average of the logarithms of the factors by which our wealth changes on a daily basis. This notion is discussed further in Section~\ref{subsection-universalization-defined}.} of such a strategy approaches that of the $\CRP$ strategy operating with the optimal, return-maximizing parameters chosen with hindsight.

This paper generalizes previous results and introduces a framework
that allows universalizations of other parameterized investment
strategies. As we see in Section~\ref{section-strategies}, investment strategies fall under two
categories; \emph{trading strategies} operate on a single stock and
dictate when to buy and short\footnote{A short position in a stock, discussed in Section~\ref{subsection-trading-strategies}, allows us to earn a profit when the stock declines in value.} the stock; \emph{portfolio strategies},
such as $\CRP$, operate on the stock market as a whole and dictate how to
allocate wealth among multiple stocks. We present several examples of 
common trading and portfolio strategies that can be universalized
in our framework. We discuss our universalization framework in Section~\ref{section-universalization}. The proofs of our results are very general and, as with previous universal portfolio results, we make no
assumptions on the underlying distribution of the stock prices; our
 results are applicable for all sequences of stock returns and
market conditions. The running times of universalization algorithms are, in
general, exponential in the number of parameters used by the
underlying investment strategy. Kalai and Vempala \cite{Kalai:2000:EAU}
presented an efficient implementation of the $\CRP$ algorithm that runs in time
polynomial in the number of parameters. In Section~\ref{section-computation}, we present general conditions
on investment strategies under which the universalization algorithm can be efficiently implemented. We also give some investment strategies that
satisfy these conditions. Section~\ref{section-further-research} concludes with directions for further research.

\section{Types of Investment Strategies}
\label{section-strategies}

Suppose we would like to distribute our wealth among $m$ stocks\footnote{We use the term ``stocks'' in order to keep our terminology consistent with previous work, but we actually mean a broader range of investment instruments, including both long and short positions in stocks.}. \emph{Investment strategies} are general classes of rules that dictate how to invest capital. At time $t>0$, a strategy $S$ takes as input an \emph{environment vector} $\env_t$ and a \emph{parameter vector} $\vecw$, and returns an \emph{investment description} $S_t(\vecw)$ specifying how to allocate our capital at time $t$. The environment vector $\env_t$ contains historic market information, including stock price history, trading volumes, \etc{}; the parameter vector $\vecw$ is independent of $\env_t$ and specifies exactly how the strategy $S$ should operate; the investment description $S_t(\vecw) = (S_{t1}(\vecw), \ldots, S_{tm}(\vecw))$ is a vector specifying the proportion of wealth to put in each stock, where we put a fraction $S_{ti}(\vecw)$ of our holdings in stock $i$, for $1\le i \le m$. For example, $\CRP$ is an investment strategy; coupled with a portfolio vector $\vecw$ it tells us to ``rebalance our portfolio on a daily basis according to $\vecw$''; its investment description, $\CRP_t(\vecw) = \vecw$, is independent of the market environment $\env_t$.

There are two types of investment strategies. \emph{Trading
strategies} tell us whether we should take a \emph{long} (bet that the
stock price will rise) or a \emph{short} (bet that the stock price
will fall) position on a given stock.  \emph{Portfolio strategies}
tell us how to distribute our wealth among various stocks. Trading
strategies are denoted by $T$, and portfolio strategies are denoted by $P$. We  use $S$ to denote either kind of
strategy.
For $k\ge 2$, let
\begin{equation}
\label{equation-weights}
\calW_k = \Set{\vecw = (w_1, \ldots, w_k) \in[0,1]^k}{\sum_{i=1}^kw_i = 1}.
\end{equation}
\begin{remark}
\label{remark-W_k}
$\calW_k$ is a $(k-1)$-dimensional simplex in $\R^k$. The investment strategies that we describe below are parameterized by vectors in $\calW_k^\ell = \calW_k\times\cdots\times\calW_k$ ($\ell$ times) for some  $k\ge 2$ and $\ell \ge 1$. We may write $\vecw \in \calW_k^\ell$ in the form $\vecw = (\vecw_1, \ldots, \vecw_\ell)$, where $\vecw_\iota = (w_{\iota 1}, \ldots, w_{\iota k})$ for $1\le \iota \le \ell$.
\end{remark}

\subsection{Trading Strategies}
\label{subsection-trading-strategies}

Suppose that our market contains a single stock. We have $m=2$
potential investments: either a \emph{long position} or \emph{short
position} in the stock. To take a \emph{long position}, we buy shares
in hopes that the share price will rise. We \emph{close a long
position} by selling the shares. The money we use to buy the shares is
our \emph{investment in the long position}; the \emph{value} of the
investment is the money we get when we close the position. If we let
$p_t$ denote the stock price at the beginning of day $t$, the value of
our investment will change by a factor of $x_{t} =
\frac{p_{t+1}}{p_t}$ from day $t$ to $t+1$.

To take a \emph{short position}, we borrow shares from our broker and sell them on the market in hopes that the share price will fall. We \emph{close a short position} by buying the shares back and returning them to our broker. As collateral for the borrowed shares, our broker has a \emph{margin requirement}: a fraction $\alpha$ of the value of the borrowed shares must be deposited in a \emph{margin account}. Should the price of the security rise sufficiently, the collateral in our margin account will not be enough, and the broker will issue a \emph{margin call}, requiring us to deposit more collateral. The margin requirement is our \emph{investment in the short position}; the \emph{value} of the investment is the money we get when we close the position.

\begin{lemma}
\label{lemma-margin}
Let the margin requirement for a short position be $\alpha \in (0,1]$. Suppose that a short position is opened on day $t$ and that the price of the underlying stock changes by a factor of $x_t = \frac{p_{t+1}}{p_t} < 1 + \alpha$ during the day. Then the value of our investment in the short position changes by a factor of $x'_t = 1 + \frac{1 - x_t}{\alpha}$ during the day.
\end{lemma}

\begin{proof}
Suppose that we have \$$v$ to deposit in the margin account. Using this as our investment in the short position, we can sell \$$v/\alpha$ worth of shares. Combining the proceeds of the stock sale with our margin account balance, we will have a total of $v + v/\alpha$ dollars. At the end of the day, it will cost $x_t v/\alpha$ dollars to buy the shares back, and we will be left with $v + \frac{v}{\alpha} - x_t \frac{v}{\alpha}$ dollars, which is positive since $x_t < 1 + \alpha$. Thus, our investment of \$$v$ in the short position has changed by a factor of $1 + \frac{1 - x_t}{\alpha}$, as claimed.
\end{proof}

Should the price of the underlying stock change by a factor greater than $1 + \alpha$, we will lose more money than we initially put in. We will assume that the margin requirement $\alpha$ is sufficiently large that the daily price change of the stock is always less than $1+\alpha$.
\begin{remark}
This assumption can be eliminated by purchasing a \emph{call option} on the stock with some strike price $p < (1+\alpha)p_t$. Should the stock price get too high, the call allows us to purchase the stock back for \$$p$. Though its price detracts from the performance of our short trading strategy, the call protects us from potentially unlimited losses due to rising stock price.
\end{remark}
If a short position is held for several days, assume that it is \emph{rebalanced} at the beginning of each day: either part of the short is closed (if $x_t > 1$) or additional shares are shorted (if $x_t < 1$) so that the collateral in the margin account is exactly an $\alpha$ fraction of the value of the shorted shares. This ensures that the value of a short position changes by a factor, $x'_t = 1 + \frac{1 - x_t}{\alpha}$, each day. Treating short positions in this 
way, they can simply be viewed as any other stock, so trading strategies are effectively investment strategies that decide between two potential investments: a long or a short position in a given stock. The investment description of a trading strategy $T$ is $T_t = (T_{t1}, T_{t2})$, where $T_{t1}$ and $T_{t2}$ are the fraction of wealth to put in a long and short position respectively. 

\begin{remark}
Let $D = T_{t1} - T_{t2}/\alpha$ be the \emph{net long position} of the investment description. In practice, if $D>0$, investors should put a $D$ fraction of their money in the long position and a $1 - D$ fraction in cash; if $D < 0$, investors should invest $D$ in the short position and $1-D$ in cash; if $D=0$, investors should avoid the stock completely and keep all their money in cash. From a practical standpoint, it is desirable for the trading strategy to be \emph{decisive}, \ie{} $|D| = 1$, so that our allocation of money to the stock is always fully invested in the stock (either as a long or a short position). We show in Section~\ref{section-universalization} that investment strategies that are continuous in their parameter spaces are universalizable. Though decisive trading strategies $T$ are discontinuous, the can be approximated by continuous startegies whose investment descriptions converge almost everywhere to $T_t$ as $t \rightarrow \infty$ (see, for example, (\ref{equation-MA-step-approx}) below).
\end{remark}

We now describe some commonly used and researched
 trading strategies \cite{Sullivan:1999:DST,Brock:1992:STT,Gartley:1935:PSM,Wyckoff:1910:STR} and show how they can be parameterized.

\paragraph{$\MA[k]$: Moving Average Cross-over with $k$-day Memory.}
In traditional applications \cite{Gartley:1935:PSM} of this rule, we compare
the current stock price with the moving average over, say, the
previous 200 days: if the price is above the moving average, we take a
long position, otherwise we take a short position. Some
generalizations of this rule have been made, where we compare a fast
moving average (over, for example, the past five to 20 days) with a
slow moving average (over the past 50 to 200 days). We generalize this
rule further. Given day $t\ge 0$, let $\vecv_t =
(v_{t1}, \ldots, v_{tk})$ be the
\emph{price-history vector} over the previous $k$ days, where
$v_{tj}$ is the stock price on day $t-j$. Assume that
the stock prices have been normalized such that $0<
v_{tj} \le 1$. Let $(\vecw_{F}, \vecw_{S}) \in
\calW_k^2$ (where $\calW_k$ is defined in
(\ref{equation-weights})) be the weights to compute the fast moving
and slow moving averages, so these averages on day $t$ are given by
$\vecw_F \cdot \vecv_t$ and $\vecw_S \cdot \vecv_t$
respectively. Since the prices have been normalized to the interval
$(0,1]$, $-1 \le (\vecw_F - \vecw_S)\cdot
\vecv_t \le 1$.  Let $g : [-1, 1] \rightarrow [0,1]$ be the
\emph{long/short allocation function}. The idea is that
$g((\vecw_F - \vecw_S)\cdot \vecv_t)$ represents the
proportion of wealth that we invest in a long position. The full
investment description for the $\MA = \MA[k]$ trading strategy is
$$
\MA_t(\vecw_F, \vecw_S) = \big(g((\vecw_F -
\vecw_S)\cdot \vecv_t), \ 1 - g((\vecw_F -
\vecw_S)\cdot\vecv_t\big).
$$ 
Note that the dimension of the
parameter space for $\MA[k]$ is $2(k-1)$ since each of $\vecw_F$
and $\vecw_S$ are taken from $(k-1)$-dimensional spaces. Possible
functions for $g$ include
\begin{alignat}{2}
\label{equation-MA-step}
g_s(x) &=
\begin{cases}
0 & \text{if $x < 0$} \\
1  & \text{otherwise}
\end{cases}
& \qquad & \text{(step function);} \\
\label{equation-MA-step-approx}
g_{(t)}(x) &=
\begin{cases}
0 & \text{if $x < - \frac{1}{t}$} \\
\frac{t}{2}(x + \frac{1}{t})  & \text{if $-\frac{1}{t} \le x \le \frac{1}{t}$} \\
1  & \text{if $\frac{1}{t} < x$}
\end{cases}
& \qquad & \text{(linear step approximation);}
\end{alignat}
and the line
\begin{equation}
\label{equation-MA-line}
g_\ell(x) = \frac{x+1}{2}
\end{equation}
that intersects $g_s(x)$ at the extreme points $x = \pm 1$ of its domain. Note that $g_{(t)}(x)$ is parameterized by the day $t$ during which it is called and that it converges to $g_s(x)$ on $[-1,1]\setminus\set{0}$ as $t$ increases.

\begin{remark}
\label{remark-cts-approx}
The long/short allocation function used in traditional applications of this rule is the step function $g_s(\cdot)$. As we see in Section~\ref{section-universalization}, in order for an investment strategy to be universalizable, its allocation function must be continuous, necessitating the continuous approximation $g_{(t)}(\cdot)$. The linear approximation $g_\ell(\cdot)$ can be used with the results of Section~\ref{section-computation}, to allow for efficient computation of the universalization algorithm.
\end{remark}

\paragraph{$\SR[k]$: Support and Resistance Breakout with $k$-day
Memory.} Discussed as early as Wyckoff \cite{Wyckoff:1910:STR} in 1910, this strategy
uses the idea that the stock price trades in a \emph{range} bounded by
\emph{support} and \emph{resistance} levels. Should the price fall
below the support level, the idea is that it will continue to fall and
a short position should be taken in the stock. Similarly, should the
price rise above the resistance level, the idea is that it will
continue to rise and a long position should be taken in the stock. If
the stock price remains between the support and resistance levels, the
idea is that it will continue to trade in this range in an
unpredictable pattern and the stock should be avoided.  Support and
resistance levels are defined quite arbitrarily in practice, usually
the minimum and maximum prices over the past $k$ days, where $k$ is
usually taken to be 50, 150, or 200 \cite{Brock:1992:STT}. To generalize this
rule, given day $t \ge 0$, let $\underline\vecv_t =
(\underline v_{t1}, \ldots,
\underline v_{tk})$ and $\overline\vecv_t =
(\overline v_{t1}, \ldots,
\overline v_{tk})$ be the minimum and maximum price
histories, where $\underline v_{tj}$ and
$\overline v_{tj}$ are the minimum and maximum prices
over the previous $j$ days, normalized so that they are in the range
$(0,1]$. Let $\vecw \in \calW_k$ be the weights to compute the
support and resistance levels, so these levels on day $t$ are given by
$s_t = \vecw \cdot \underline\vecv_t$ and $r_t = \vecw \cdot
\overline\vecv_t$ respectively.

\begin{lemma}
\label{lemma-s-bigger-than-r}
The support level is bounded above by the resistance level: $s_t \le r_t$.
\end{lemma}

\begin{proof}
This follows from the fact that for all $1\le j\le k$, $\underline v_{tj} \le \overline v_{tj}$.
\end{proof}

The long/short allocation function will be denoted by $h : \Set{(x,y)\in[-1, 1]^2}{x\le y} \rightarrow [0,1]$. Let $p_t$ be the current stock price (normalized to $(0,1]$ along with $\underline\vecv_t$ and $\overline\vecv_t$). The idea is that $h(p_t - r_t, p_t - s_t)$ tells us the proportion of wealth that we invest in a long position. The full investment description for the $\SR = \SR[k]$ trading strategy is 
$$
\SR_t(\vecw) = \big(h(p_t - r_t, p_t - s_t), \ 1 - h(p_t - r_t, p_t - s_t)\big).
$$ 
The value of $h$ need only be defined on $\Set{(x,y)\in[-1, 1]^2}{x\le y}$ since, by Lemma~\ref{lemma-s-bigger-than-r}, $s_t \le r_t$. A possible function for $h$ is
\begin{equation}
\label{equation-SR-step}
h_s(x,y) =
\begin{cases}
0   & \text{if $x \le y \le 0$} \\
\frac{1}{\alpha+1} & \text{if $x < 0 < y$} \\
1   & \text{if $y \ge x \ge 0$} \\
\end{cases}
\qquad\text{(step function),}
\end{equation}
where the investment allocation $\frac{1}{\alpha+1}$ long, $1 - \frac{1}{\alpha+1} = \frac{\alpha}{\alpha+1}$ short is equivalent to having no position in the stock, since the return from such an allocation is $\frac{x_t}{\alpha+1} + (1 + \frac{1 - x_t}{\alpha})\frac{\alpha}{\alpha+1} = 1$. Other possibilities include a continuous approximation \refstepcounter{equation} \label{equation-SR-step-approx} $h_{(t)}(x,y)$ to $h_s(x,y)$ with maximum slope at most $\frac{1}{t}$ (defined similarly to $g_{(t)}(x)$) (\theequation), or the plane
\begin{equation}
\label{equation-SR-plane}
h_p(x,y) = \frac{(x+1)\alpha}{2(\alpha+1)} + \frac{y+1}{2(\alpha+1)}
\end{equation}
that intersects $h_s(x,y)$ at the extreme points $(x,y) = (-1, -1)$, $(-1, 1)$, and $(1, 1)$ of its domain.

\subsection{Portfolio Strategies}

Portfolio strategies are investment strategies that distribute wealth among $m$ stocks. The investment description of a portfolio strategy $P$ is $P_t = (P_{t1}, \ldots, P_{tm})$, where $0\le P_{ti} \le 1$ and $\sum_{i=1}^mP_{ti} = 1$. We put a fraction $P_{ti}$ of our wealth in stock $i$ at time $t$. 

\paragraph{${\mathbf \CRP}$: Constantly Rebalanced Portfolio
\cite{Cover:1991:UP}.} The parameter space for the $\CRP$ strategy is $\bbW
= \calW_m$. The investment description is $\CRP_t(\vecw) =
\vecw$: at the beginning of each day, we invest a $w_i$
proportion of our wealth in stock $i$.

\paragraph{${\mathbf \CRPside}$: Constantly Rebalanced Portfolio with Side Information.} Cover and Ordentlich \cite{Cover:1996:UPS} consider a generalization of $\CRP$. Rather than rebalancing our holdings according to a single portfolio vector $\vecw \in \calW_m$ every day, we have $k$ vectors $\vecw_1, \ldots, \vecw_k \in \calW_m$ and a side information state $y_t \in \set{1, \ldots, k}$ that classifies each day $t$ into one of $k$ possible categories; on day $t$ we rebalance our holdings according to $\vecw_{y_t}$. By partitioning the time interval into $k$ subsequences corresponding to each of the $k$ side information states and running $k$ instances of the universalization algorithm (one instance for each state), Cover and Ordentlich show that the average daily return approaches that of the underlying strategy operating with $k$ optimal parameters, $\vecw^*_1, \ldots, \vecw^*_k \in \calW_m$, where $\vecw^*_j$ is used on days $t$ when the side information state is $y_t = j$. We generalize this further by allowing portions of our wealth to be rebalanced according to several of the $\vecw_j$ every day. Suppose that the side information is encapsulated in some vector $\vecv\in \R^\ell$, for some $\ell$. This vector can contain information about specific stocks, such as historic
performance and company fundamentals, or macro-economic indicators
such as inflation and unemployment. Let $\vecf = (f_1, \ldots, f_k) :
\R^\ell \rightarrow [0,1]^k$ be some function satisfying $\sum_{j=1}^k
f_j(\vecv) = 1$ for all $\vecv\in\R^\ell$. The parameter space is $\calW_m^k$; the investment
description is $\CRPside_t(\vecw_1, \ldots, \vecw_k) =
\sum_{j=1}^k f_j(\vecv_t)\vecw_j$, where $\vecv_t$ is the
indicator vector for day $t$. Under such a scheme, we have the
flexibility of splitting our wealth among multiple sets of portfolios
 $\vecw_1, \ldots, \vecw_k$ on any given day, rather than
being forced to choose a single one. For example, assume that $\vecv$ is a $k$-dimensional vector, with each $v_i$ corresponding to portfolio $\vecw_i$. Define $\vecf : \R^k \rightarrow [0,1]^k$ by $f_i(\vecv_t) = \frac{v_{ti}}{\sum_{\iota = 1}^k v_{t\iota}}$, so that our allocation is biased towards portfolios corresponding to higher indicators while still maintaining a position in the others.

\paragraph{$\IA[k]$: $k$-Way Indicator Aggregation.} For each day
$t\ge 0$, suppose that each stock $i$ has a set of $k$ indicators
$\vecv_{ti} = (v_{ti1}, \ldots,
v_{tik})$, where each $v_{tij} \in
(0,1]$ and, for $1\le j\le k$, $v_{t1j}, \ldots,
v_{tmj}$ have been normalized such that there is at
least one $i$ such that $v_{tij} = 1$. Examples of
possible indicators include historic stock performance and trading
volumes, and company fundamentals. Our goal is to aggregate the
indicators for each stock to get a measure of the stock's
attractiveness and put a greater proportion of our wealth in stocks
that are more attractive. We will aggregate the indicators by taking
their weighted average, where the weights will be determined by the
parameters. The parameter space is $\bbW = \calW_k$ and the
investment description is 
$$
\IA_t(\vecw) =
\big( \textstyle\frac{\vecw\cdot\vecv_{t1}}{\sum_{i=1}^m
 \vecw\cdot\vecv_{ti}}, \ldots,
\textstyle\frac{\vecw\cdot\vecv_{tm}}{\sum_{i=1}^m
\vecw\cdot\vecv_{ti}}\big).
$$

\section{Universalization of Investment Strategies}
\label{section-universalization}

\subsection{Universalization Defined}
\label{subsection-universalization-defined}

In a typical stock market, wealth grows geometrically. On day $t\ge 0$, let $\vecx_t$ be the \emph{return vector} for day $t$, the vector of factors by which stock prices change on day $t$. The return vector corresponding to a trading strategy on a single stock is $(x_t, 1 + \frac{1 - x_t}{\alpha})$, where $x_t$ is the factor by which the price of the stock changes and $1 + \frac{1 - x_t}{\alpha}$ is the factor by which our investment in a short position changes, as described in Lemma~\ref{lemma-margin}; the return vector corresponding to a portfolio strategy is $(x_{t1}, \ldots, x_{tm})$, where $x_{ti}$ is the factor by which the price of stock $i$ changes, where $1\le i\le m$. Henceforth, we do not make a distinction between return vectors corresponding to trading and portfolio strategies; we assume that $\vecx_t$ is appropriately defined to correspond to the investment strategy in question. For an investment strategy $S$ with parameter vector $\vecw$, the \emph{return of $S(\vecw)$} during the $t$-th day---the factor by which our wealth changes on the $t$-th day when invested according to $S(\vecw)$---is $S_t(\vecw)\cdot\vecx_t = \sum_{i=1}^m S_{ti}(\vecw)\cdot x_{ti}$ (recall that $S_t(\vecw)$ is the investment description of $S(\vecw)$ for day $t$, which is a vector specifying the proportion of wealth to put in each stock). Given time $n>0$, let $\Ret_n(S(\vecw)) = \prod_{t=0}^{n-1}S_t(\vecw)\cdot\vecx_t$ be the \emph{cumulative return of $S(\vecw)$ up to time $n$}; we may write $\Ret_n(\vecw)$ in place of $\Ret_n(S(\vecw))$ if $S$ is obvious from context. We analyze the performance of $S$ in terms of the \emph{normalized log-return} $\LogRet_n(\vecw) = \LogRet_n(S(\vecw)) = \frac{1}{n} \log \Ret_n(\vecw)$ of the wealth achieved.

For investment strategy $S$, let $\vecw_n^* = \arg\max_{\vecw\in\R^*} \Ret_n(S(\vecw))$ be the parameters that maximize the return of $S$ up to day $n$.\footnote{As mentioned above, $\vecw_n^*$ can only be computed with hindsight.} An investment strategy $U$ \emph{universalizes} (or \emph{is universal for}) $S$ if\footnote{Unlike previously discussed investment strategies, the behavior of $U$ is fully defined without an additional parameter vector $\vecw$.}
$$
\textstyle
\LogRet_{n}(U) = \LogRet_{n}(S(\vecw_{n}^*)) - \oh(1)
$$
for all environment vectors $\env_n$. That is, $U$ is universal for $S$ if the average daily log-return of $U$ approaches the optimal average daily log-return of $S$ as the length $n$ of the time horizon grows, regardless of stock price sequences.

\subsection{General Techniques for Universalization}
Given an investment strategy $S$, let $\bbW$ be the parameter space for $S$ and let $\mu$ be the uniform measure over $\bbW$. Our universalization algorithm for $S$, $\U(S)$, is a generalization of Cover's original result \cite{Cover:1991:UP}. The investment description $\U_t(S)$ for the universalization of $S$ on day $t > 0$ is a weighted average of the $S_t(\vecw)$ over $\vecw\in\bbW$, with greater weight given to parameters $\vecw$ that have performed better in the past (\ie{} $\Ret_t(\vecw)$ is larger). Formally, the investment description is
\begin{equation}
\label{equation-universal}
\U_t(S) 
= \frac{\int_\bbW S_{t}(\vecw) \Ret_{t}(\vecw) d\mu(\vecw)} {\int_\bbW \Ret_{t}(\vecw) d\mu(\vecw)}
= \frac{\int_\bbW S_{t}(\vecw) \Ret_{t}(S(\vecw)) d\mu(\vecw)} {\int_\bbW \Ret_{t}(S(\vecw)) d\mu(\vecw)},
\end{equation}
where we take $\Ret_{0}(\vecw) = 1$ for all $\vecw\in\bbW$.\footnote{Cover's algorithm is a special case of this, replacing $S_t(\vecw)$ with $\vecw$.}
\begin{remark}
The definition of universalization can be expanded to include measures other than $\mu$, but we consider only $\mu$ in our results.
\end{remark}

\begin{lemma}[\cite{Blum:1999:UPT,Cover:1996:UPS}]
\label{lemma-average}
The cumulative $n$-day return of $\U(S)$ is 
$$
\Ret_{n}(\U(S)) = \int_\bbW \Ret_{n}(\vecw) d\mu(\vecw) = \Expect\big(\Ret_{n}(\vecw)\big),
$$ the $\mu$-weighted average of the cumulative returns of the investment strategies $\Set{S(\vecw)}{\vecw \in \bbW}$.
\end{lemma}

\begin{proof}
The return of $\U(S)$ on day $t$ is $\U_t(S) \cdot \vecx_{t}$, where $\vecx_{t}$ is the return vector for day $t$. The cumulative $n$-day return of $\U(S)$ is
\begin{eqnarray*}
\Ret_{n}(\U(S)) &=& \prod_{t=0}^{n-1} \U_t(S) \cdot \vecx_{t}
= \prod_{t=0}^{n-1} \frac{\int_\bbW S_{t}(\vecw) \Ret_{t}(\vecw) d\mu(\vecw)} {\int_\bbW \Ret_{t}(\vecw) d\mu(\vecw)} \cdot \vecx_{t} \\
&=& \prod_{t=0}^{n-1} \frac{\int_\bbW (S_{t}(\vecw) \cdot \vecx_{t}) \Ret_{t}(\vecw) d\mu(\vecw)} {\int_\bbW \Ret_{t}(\vecw) d\mu(\vecw)}
= \prod_{t=0}^{n-1} \frac{\int_\bbW \Ret_{t+1}(\vecw) d\mu(\vecw)} {\int_\bbW \Ret_{t}(\vecw) d\mu(\vecw)}.
\end{eqnarray*}
The result follows from the fact that this product telescopes.
\end{proof}

Rather than directly universalizing a given investment strategy $S$, we instead focus on a modified version of $S$ that puts a nonzero fraction of wealth in each of the $m$ stocks. Define the investment strategy $\bar S$ by 
$$
\bar S_t(\vecw) = (1 - \frac{\varepsilon}{2(t+1)^2})S_t(\vecw) + \frac{\varepsilon}{2m(t+1)^2}
$$ 
for $t \ge 0$ and some fixed $0 < \varepsilon < 1$. Rather than universalizing $S$, we instead universalize $\bar S$. Lemma~\ref{lemma-modified-strategy} tells us that we do not lose much by doing this.

\begin{lemma}
\label{lemma-modified-strategy}
For all $n \ge 0$, (1) $\Ret_n(\U(\bar S)) \ge (1 - \varepsilon)\Ret_n(\U(S))$ and (2) $\LogRet_n(\U(\bar S)) = \LogRet_n(\U(S)) - \frac{\oh(n)}{n}$. (3) If $\U(S)$ is a universalization of $S$, then $\U(\bar S)$ is a universalization of $S$ as well.
\end{lemma}
\begin{proof}
Statements (2) and (3) follow directly from (1). Statement (1) follows from the fact that for all $\vecw\in\bbW$,
$\Ret_n(\bar S(\vecw)) = \prod_{t=0}^{n-1}\bar S_t(\vecw)\cdot\vecx_t
\ge \prod_{t=0}^{n-1}(1 - \frac{\varepsilon}{2(t+1)^2})S_t(\vecw)\cdot\vecx_t
\ge (1 - \sum_{t=0}^{n-1}\frac{\varepsilon}{2(t+1)^2})\Ret_n(S(\vecw))
\ge (1 - \varepsilon)\Ret_n(S(\vecw))$.
\end{proof}

\begin{remark}
\label{remark-S-bdd-below}
Henceforth, we assume that suitable modifications have been made to $S$ to ensure that $S_{ti}(\vecw) \ge \frac{\varepsilon}{2m(t+1)^2}$ for all $1\le i\le m$ and $t \ge 0$.
\end{remark}

\begin{theorem}
\label{theorem-main-universalization}
Given an investment strategy $S$, let $\bbW = \calW_k^{\ell}$ (for some $k \ge 2$ and $\ell \ge 1$) be its parameter space.  For $1\le i\le m$, $1\le\iota \le \ell$ and $1\le j\le k$, assume
that there is a constant $c$ such that $\abs{\frac{\partial
S_{ti}(\vecw)}{\partial w_{\iota j}}} \le c(t+1)$ for all
$\vecw\in\bbW$. Then $\U(S)$ is a universalization of $S$.
\suppress{
$\inc:\R \rightarrow \R^+$ any unbounded monotone
increasing function and let $\eff(t) = \lceil\frac{n}{\inc(n) \log
(n+1)}\rceil$.  Let $\ell_t = (k-1)\ell\eff(t)$ be the dimension of
$\bbW^{\eff(t)}$. For $1\le i\le m$ and $1\le j\le \ell_t$, assume
that there is a constant $c$ such that $\abs{\frac{\partial
S_{ti}(\vecw)}{\partial w_j}} \le c(t+1)$ for all
$\vecw\in\bbW^{\eff(t)}$. Then $\D(S) = \D(S, \eff)$
is a dynamic $\eff$-universalization of $S$.}
\end{theorem}

To prove Theorem~\ref{theorem-main-universalization}, we first prove some preliminary results.
\begin{lemma}
\label{lemma-vector-ratio}
For nonnegative vector $\veca$ and strictly positive vectors $\vecb$ and $\vecx$,
$$
\min_{i}\frac{a_i}{b_i} \le \frac{\veca\cdot\vecx}{\vecb\cdot\vecx} \le \max_i\frac{a_i}{b_i}.
$$
\end{lemma}
\begin{proof}
Assume that the components of $\veca$ and $\vecb$ are strictly positive. Otherwise, the lemma holds trivially. Let $i_{\max} = \arg\max_i\frac{a_i}{b_i}$ and $i_{\min} = \arg\min_i\frac{a_i}{b_i}$, so that 
$$
\frac{a_i}{b_i} \le \frac{a_{i_{\max}}}{b_{i_{\max}}} \Leftrightarrow \frac{a_i}{a_{i_{\max}}} \le \frac{b_i}{b_{i_{\max}}}
\quad \text{and} \quad
\frac{a_i}{b_i} \ge \frac{a_{i_{\min}}}{b_{i_{\min}}} \Leftrightarrow \frac{a_i}{a_{i_{\min}}} \ge \frac{b_i}{b_{i_{\min}}}.
$$
Then
$$ \begin{array}{crcccl}
&
\displaystyle\frac{a_{i_{\min}}(x_{i_{\min}} + \sum_{i\not=i_{\min}}\frac{a_i}{a_{i_{\min}}}x_i)}{b_{i_{\min}}(x_{i_{\min}} + \sum_{i\not=i_{\min}}\frac{b_i}{b_{i_{\min}}}x_i)}
&=& \displaystyle\frac{\veca\cdot\vecx}{\vecb\cdot\vecx}
&=& \displaystyle\frac{a_{i_{\max}}(x_{i_{\max}} + \sum_{i\not=i_{\max}}\frac{a_i}{a_{i_{\max}}}x_i)}{b_{i_{\max}}(x_{i_{\max}} + \sum_{i\not=i_{\max}}\frac{b_i}{b_{i_{\max}}}x_i)} \\[20pt]
\Rightarrow&
\displaystyle\frac{a_{i_{\min}}}{b_{i_{\min}}}
&\le& \displaystyle\frac{\veca\cdot\vecx}{\vecb\cdot\vecx}
&\le& \displaystyle\frac{a_{i_{\max}}}{b_{i_{\max}}}.
\end{array} $$
\end{proof}

Our next two results are related to the $(k-1)$-dimensional volumes of some subsets of $\R^k$.

\begin{lemma}
\label{lemma-volume-simplex}
The $(k-1)$-dimensional volume of the simplex $\calW_k = \Set{\vecw \in [0,1]^k}{\sum_{i=1}^{k}w_i = 1}$, defined in (\ref{equation-weights}), is $\frac{\sqrt{k}}{(k-1)!}$.
\end{lemma}
\begin{proof}
By induction on $k$, it can be shown that the $k$-dimensional volume of the solid $W_k(s) = \Set{\vecw}{\sum_{i=1}^{k}w_i \le s}$ is $\frac{s^k}{k!}$. Written in terms of the length $r$ of the line segment passing between the origin and $(\frac{s}{k}, \ldots, \frac{s}{k}) \in \R^k$, the volume is $\frac{1}{k!}r^k k^{\frac{k}{2}}$ since $s = r\sqrt{k}$. Upon differentiation with respect to $r$, $\frac{1}{(k-1)!}r^{k-1} k^{\frac{k}{2}} = \frac{1}{(k-1)!}\sqrt{k}s^{k-1}$, we arrive at the $(k-1)$-dimensional volume of the simplex $\calW_k(s) = \Set{\vecw}{\sum_{i=1}^{k}w_i = s}$. Setting $s=1$ yields the desired result.
\end{proof}

\begin{lemma}
\label{lemma-volume-ball}
The $(k-1)$-dimensional volume of a $(k-1)$-dimensional ball of radius $\rho$ embedded in $\calW_k$ is $\frac{\pi^{\frac{k-1}{2}}\rho^{k-1}}{\Gamma(\frac{k-1}{2} + 1)}$, where
$$
\textstyle{\Gamma(\ell) = (\ell-1)! \quad\text{and}\quad \Gamma(\ell+\frac{1}{2}) = (\ell - \frac{1}{2})(\ell - \frac{3}{2})\cdots(\frac{1}{2})\sqrt{\pi}}.
$$
\end{lemma}

\begin{proof}
This result is proven in Folland \cite[Corollary 2.56]{Folland:1984:RAM}.
\end{proof}

\begin{farproof}{Theorem~\ref{theorem-main-universalization}}
>From Lemma~\ref{lemma-average}, the return of $\U(S)$ is the average of the cumulative returns of the investment strategies $\Set{S(\vecw)}{\vecw \in \bbW}$. Let $\vecw^* = \arg\max_{\vecw\in\bbW}\Ret_n(S(\vecw))$ be the parameters that maximize the return of $S$. We show that there is a set $B$ of nonzero volume around $\vecw^*$ such that for $\vecw\in B$, the return $\Ret_n(\vecw)$ is close to the optimal return $\Ret_n(\vecw^*)$. We then show that the contribution to the average return from $B$ is sufficiently large to ensure universalizability. We begin by bounding the magnitude of the gradient vector $\nabla\Ret_n(\vecw)$. From Remark~\ref{remark-S-bdd-below} and our assumption in the statement of the theorem, for all $\vecw$, $t$, $i$, $\iota$, and $j$
 $$\frac{\abs{\frac{\partial S_{ti}(\vecw)}{\partial w_{\iota j}}}}{S_{ti}(\vecw)} \le c'm(t+1)^3,$$
where $c' = \frac{2c}{\varepsilon}$. 
 Using this fact and Lemma~\ref{lemma-vector-ratio}, the partial derivative of the return function $\Ret_n(\vecw) = \Ret_n(S(\vecw)) = \prod_{t=0}^{n-1} r_t(S(\vecw))$ with respect to parameter $w_{\iota j}$ is
\begin{eqnarray*}
\abs{\frac{\partial\Ret_n(\vecw)}{\partial w_{\iota j}}}
&\le& \Ret_n(\vecw)\sum_{t=0}^{n-1}\frac{\abs{\frac{\partial(S_t(\vecw)\cdot\vecx_t)}{\partial w_{\iota j}}}}{S_t(\vecw)\cdot\vecx_t}
\le \Ret_n(\vecw)\sum_{t=0}^{n-1}\frac{\sum_{i=1}^m \abs{\frac{\partial S_{ti}(\vecw)}{\partial w_{\iota j}}}\cdot x_{ti}}{\sum_{i=1}^m  S_{ti}(\vecw)\cdot x_{ti}} \\
&\le& \Ret_n(\vecw) \sum_{t=0}^{n-1} c'm(t+1)^3 
\le c'\Ret_n(\vecw)mn^4
\end{eqnarray*}
and 
\begin{equation}
\label{equation-gradient}
|\nabla\Ret_n(\vecw)| \le c'\Ret_n(\vecw)mn^4\sqrt{k\ell}.
\end{equation}
We would like to take our set $B$ to be some $d$-dimensional ball around $\vecw^*$; unfortunately, if $\vecw^*$ is on (or close to) an edge of $\bbW$, the reasoning introduced at the beginning of this proof is not valid. We instead perturb $\vecw^*$ to a point $\tilde\vecw$ that is at least 
$$
\rho = \frac{\gamma}{c'mn^4 k^2\ell}
$$ 
away from all edges, where $0 < \gamma < 1$ is a constant, and such that $\Ret_n(\tilde\vecw)$ is close to $\Ret_n(\vecw^*)$. To illustrate the perturbation, let $\vecw^* = (\vecw^*_1, \ldots, \vecw^*_\ell)$ where
$
\vecw^*_\iota = (w^*_{\iota 1}, \ldots, w^*_{\iota k})
$
and $w^*_{\iota k} = 1 - \sum_{i=1}^{k-1}w^*_{\iota i}$ for $1 \le \iota\le \ell$. We perturb each $\vecw^*_{\iota}$ in the same way. Let $\tilde \vecw^0_\iota = \vecw^*_\iota$. For $1 \le j \le k$, given $\tilde\vecw^{j-1}_\iota$, define $\tilde \vecw^j_\iota$ as follows. Let $j_{\max}$ be the index of the maximum coordinate of $\tilde \vecw^{j-1}_\iota$. If $0 \le \tilde w^j_{\iota j} < \rho$, define 
$\tilde w^j_{\iota j} = \tilde w_{\iota j}^{j-1} + \rho$,
$\tilde w^{j}_{\iota j_{\max}} = \tilde w_{\iota j_{\max}}^{j-1} - \rho$ 
and leave all other coordinates unchanged. Otherwise, let 
$\tilde\vecw^j_{j_0} = \tilde\vecw^{j-1}_{j_0}$. The final perturbation is $\tilde\vecw = (\tilde\vecw_1, \ldots, \tilde\vecw_\ell)$, where $\tilde\vecw_\iota = \tilde\vecw^k_{\iota}$. By construction, $\tilde \vecw \in \bbW$, $\tilde\vecw$ is at least $\rho$ away from the edges of $\bbW$ and $|w^*_{\iota j} - \tilde w_{\iota j}| \le k\rho$ for all $\iota$ and $j$. We bound $\frac{\Ret_n(\vecw^*)}{\Ret_n(\tilde \vecw)}$ by the multivariate mean value theorem and the Cauchy-Schwartz inequality:
\begin{eqnarray*}
\Ret_n(\tilde\vecw) &=& \Ret_n(\vecw^*) + \Ret_n(\tilde\vecw) - \Ret_n(\vecw^*) \\
&\ge& \Ret_n(\vecw^*) - |\nabla\Ret_n(\vecw')\cdot(\tilde\vecw - \vecw^*)| 
\quad\text{(for some $\vecw'$ between $\tilde\vecw$ and $\vecw^*$)}\\
&\ge& \Ret_n(\vecw^*) - |\nabla\Ret_n(\vecw')| \cdot |\tilde\vecw - \vecw^*|
\ge \Ret_n(\vecw^*) - c'\Ret_n(\vecw')mn^4\sqrt{k\ell} \cdot k\rho\sqrt{k\ell} \\
&\ge& \Ret_n(\vecw^*) - c'\Ret_n(\vecw^*)mn^4\sqrt{k\ell} \cdot k\rho\sqrt{k\ell}
\ge \Ret_n(\vecw^*)(1 - \gamma).
\end{eqnarray*}

For $0\le\iota\le\ell$ let $C_\iota = \Set{\vecw_\iota \in\R^{k}}{|\tilde\vecw_\iota - \vecw_\iota| \le \rho}$. From the construction of $\tilde\vecw$, $B_\iota = C_\iota\cap\calW_k$ is a $(k-1)$-dimensional ball of radius $\rho$. Let $\tilde\vecw^*_\iota = \arg\max_{\vecw \in B_\iota} \Ret_n(\vecw)$ and let $\tilde\vecw^* = (\tilde\vecw^*_1, \ldots, \tilde\vecw^*_\ell)$ be the profit maximizing parameters in $B = B_1 \times\cdots\times B_\ell$. For $\vecw \in B$,
\begin{eqnarray*}
\Ret_n(\vecw) &=& \Ret_n(\tilde \vecw^*) + \Ret_n(\vecw) - \Ret_n(\tilde \vecw^*) \\
&\ge& \Ret_n(\tilde \vecw^*) - |\nabla\Ret_n(\vecw')| \cdot |\tilde\vecw^* - \vecw| \quad \text{(for some $\vecw'$ between $\tilde\vecw^*$ and $\vecw$)} \\
&\ge& \Ret_n(\tilde \vecw^*) - c'\Ret_n(\tilde \vecw^*)mn^4\sqrt{k\ell} \cdot 2\rho\sqrt{\ell} 
\ge \Ret_n(\tilde \vecw^*)(1 - \gamma) \\
&\ge& \Ret_n(\vecw^*)(1 - 2\gamma).
\end{eqnarray*}

By Lemma~\ref{lemma-average}
\begin{eqnarray*}
\Ret_n(\U(S)) &=& \int_\bbW \Ret_{n}(S(\vecw)) d\mu(\vecw) 
\ge \int_B \Ret_{n}(\vecw) d\mu(\vecw)
\ge (1 - 2\gamma)\Ret_n(\vecw^*) \int_B d\mu(\vecw) \\
&\ge& (1 - 2\gamma)\Ret_n(\vecw^*)\frac{\int_B d\vecw}{\int_\bbW d\vecw} \\
&=& (1 - 2\gamma)\Ret_n(\vecw^*)\left(\frac{\pi^{\frac{k-1}{2}}\rho^{k-1}}{\Gamma(\frac{k-1}{2} + 1)} \cdot \frac{(k-1)!}{\sqrt{k}}\right)^\ell \quad \text{(from Lemmas~\ref{lemma-volume-simplex} and \ref{lemma-volume-ball})} \\
&=& \Ret_n(\vecw^*)\Lambda(\gamma, m, k, \ell)n^{-4k\ell}
\end{eqnarray*}
where $\Lambda$ is some constant depending on $\gamma$, $m$, $k$, and $\ell$. Therefore,
\begin{equation}
\label{equation-universal-proof}
\LogRet_n(\vecw^*) - \LogRet_n(\U(S)) 
\le \frac{\log \Lambda(\gamma, m, k, \ell)}{n} + 4k\ell\frac{\log n}{n} = \frac{\oh(n)}{n},
\end{equation}
as claimed.
\end{farproof}

\begin{remark}
The techniques used in the proof of Theorem~\ref{theorem-main-universalization} can be generalized to other investment strategies with bounded parameter spaces $\bbW$ that are not necessarily of the form $\calW_k^\ell$.
\end{remark}

\subsection{Increasing the Number of Parameters with Time}

The reader may notice from the proof of Theorem~\ref{theorem-main-universalization} that an investment strategy $S$ may be universalizable even if the dimensions of its parameter space $\bbW$ grow with time. In fact, even if the dimension of the parameter space (the coefficient of $\frac{\log n}{n}$ in (\ref{equation-universal-proof})) is $\Oh(\frac{n}{\phi(n)\log n})$, where $\phi(n)$ is a monotone increasing function, the strategy is still universalizable. This introduces an interesting possibility for investment strategies whose parameter spaces grow with time as more information becomes available. As a simple example, consider \emph{dynamic universalization}, which allows us to track a higher-return benchmark than basic universalization. Partition the time interval $\timeInt = [0,n)$ into 
$\eff = \Oh(\frac{n}{\phi(n)\log n})$
 subintervals $\timeInt_1, \ldots, \timeInt_\eff$ and let $\vecw^*_{\timeInt_j}$ be the
parameters that optimize the return during $\timeInt_j$. In $\timeInt_1$, we run the universalization algorithm given by (\ref{equation-universal}) over the basic parameter space $\bbW$ of $S$. In $\timeInt_2$, we run the algorithm over $\bbW\times\bbW$; to compute the investment description for a day $t\in\timeInt_2$ using (\ref{equation-universal}), we compute the return $\Ret_t(\vecw_1, \vecw_2)$ as the product of the returns we would have earned in $\timeInt_1$ using $\vecw_1$ and what we would have earned up to day $t$ in $\timeInt_2$ using $\vecw_2$. We proceed similarly in intervals $\timeInt_3$ through $\timeInt_\eff$. This will allow us to track the strategy that uses the optimal parameters $\vecw^*_{\timeInt_j}$ corresponding to each $\timeInt_j$. Such a strategy is useful in environments where optimal investment styles (and the optimal investment strategy parameters that go with them) change with time.

\subsection{Applications to Trading Strategies}

By proving an upper bound on $\abs{\frac{\partial T_{ti}(\vecw)}{\partial w_j}}$ for our trading strategies $T$, we show that they are universalizable.

\begin{theorem}
\label{theorem-ma}
The moving average cross-over trading strategy, $\MA[k]$, is universalizable for the long/short allocation functions $g_{(t)}(x)$ and $g_\ell(x)$ defined in (\ref{equation-MA-step-approx}) and (\ref{equation-MA-line}) respectively.
\end{theorem}

\begin{proof}
The parameters for $\MA[k]$ are of the form 
$\vecw_F = (w_{F1}, \ldots, w_{F(k-1)}, 1 - w_{F1} - \cdots - w_{F(k-1)})$ and
$\vecw_S = (w_{S1}, \ldots, w_{S(k-1)}, 1 - w_{S1} - \cdots - w_{S(k-1)})$.
Using the long/short allocation function $g_{(t)}(x)$ defined in (\ref{equation-MA-step-approx}), the partial derivative of the investment description with respect to a parameter $w_{Fj}$ (or similarly $w_{Sj}$) is
$$
\abs{\frac{\partial \MA_{ti}(\vecw_F, \vecw_S)}{\partial w_{Fj}}}
= \abs{\frac{\partial g((\vecw_F - \vecw_S)\cdot\vecv_t)}{\partial w_{Fj}}}
\le \frac{t}{2}\cdot (v_{tj} - v_{tk}) \le \frac{t}{2}
$$
where $1\le j < k$ and $i \in \set{1,2}$. Similarly, we can show that using the long/short allocation function $g_{\ell}(x)$ defined in (\ref{equation-MA-line}), $\abs{\frac{\partial \MA_{ti}(\vecw_F, \vecw_S)}{\partial w_{Fj}}} \le \frac{1}{2}$.
\end{proof}

\begin{theorem}
The support and resistance breakout trading strategy, $\SR[k]$, is universalizable for the long/short allocation functions $h_{(t)}(x,y)$ and $h_p(x,y)$ defined in (\ref{equation-SR-step-approx}) and (\ref{equation-SR-plane}) respectively.
\end{theorem}
\begin{proof}
We arrive at the result by differentiating the long/short allocation functions $h_{(t)}(x,y)$ and $h_p(x,y)$ with respect to an arbitrary parameter $w_j$ and showing that the partial derivative is $\Oh(t)$, as in the proof of Theorem~\ref{theorem-ma}.
\end{proof}

\subsection{Applications to Portfolio Strategies}

\begin{theorem}
The constantly rebalanced portfolio, $\CRP$, and $\CRP$ with side information, $\CRPside$, portfolio strategies are universalizable.
\end{theorem}
\begin{proof}
The partial derivatives of $\CRP_{ti}$ and $\CRPside_{ti}$ with respect to an arbitrary parameter $w_j$ are at most $1$.
\end{proof}

\begin{theorem}
The $k$-way indicator aggregation portfolio strategy, $\IA[k]$, is universalizable.
\end{theorem}
\begin{proof}
First, we show that $\sum_{\ell=1}^m\vecw\cdot\vecv_{t\ell} \ge \frac{1}{k}$ for all $t$. Since $\sum_{j=1}^k w_j = 1$, there exists $j_0$ such that $w_{j_0} \ge \frac{1}{k}$. Then
$\sum_{\ell=1}^m\vecw\cdot\vecv_{t\ell}
\ge \sum_{\ell=1}^mw_{j_0}\cdot v_{t\ell j_0}
\ge \frac{1}{k} \sum_{\ell=1}^m v_{t\ell j_0} \ge \frac{1}{k}$
since the $\set{v_{t\ell j_0}}_{1\le\ell\le m}$ have been normalized such that there is at least one $\ell_0$ such that $v_{t\ell_0 j_0} = 1$.

Now, let $S = \IA[k]$. By Theorem~\ref{theorem-main-universalization}, we need only show that $\frac{\partial S_{ti}(\vecw)}{\partial w_j} = \Oh(t)$, for $1\le j\le k-1$. For $t\ge 0$ and $1\le i\le m$ recall that $S_{ti}(\vecw) = \frac{\vecw\cdot\vecv_{ti}}{\sum_{\ell=1}^m\vecw\cdot\vecv_{t\ell}}$. Then, for $1\le j\le k-1$, since $\vecw = (w_1, \ldots, w_{k-1}, 1 - (w_1 +\cdots +w_{k-1}))$,
\begin{eqnarray*}
\frac{\partial S_{ti}(\vecw)}{\partial w_j}
&=& \frac{v_{tij} - v_{tik}}{\sum_{\ell=1}^m\vecw\cdot\vecv_{t\ell}} - \frac{\vecw\cdot\vecv_{ti}}{(\sum_{\ell=1}^m\vecw\cdot\vecv_{t\ell})^2}\cdot \sum_{\ell=1}^m(v_{t\ell j} - v_{t\ell k}) \\
&\le& \frac{1}{\sum_{\ell=1}^m\vecw\cdot\vecv_{t\ell}} + \frac{m}{(\sum_{\ell=1}^m\vecw\cdot\vecv_{t\ell})^2} \le k + mk^2,
\end{eqnarray*}
as we wanted to show.
\end{proof}

\section{Fast Computation of Universal Investment Strategies}
\label{section-computation}

\subsection{Approximation by Sampling}
\label{subsection-approximation}

The running time of the universalization
algorithm depends on the time to compute the integral in (\ref{equation-universal}). A
straightforward evaluation of it takes time exponential
in the number of parameters. Following Kalai and Vempala \cite{Kalai:2000:EAU},
we propose to approximate it by sampling the parameters
according to a biased distribution, giving greater
weight to better performing parameters. Define the measure $\zeta_t$ on $\bbW$ by
$$
d\zeta_t(\vecw) = \frac{\Ret_t( S(\vecw))}{\int_{\bbW}\Ret_t( S(\vecw))d\mu(\vecw)}d\mu(\vecw).
$$

\begin{lemma}[\cite{Kalai:2000:EAU}]
The investment description $\U_t(S)$ for universalization is the average of $S_t(\vecw)$ with respect to the $\zeta_t$ measure.
\end{lemma}
\begin{proof}
The average of $S_t(\vecw)$ with respect to $\zeta_t$ is
\begin{eqnarray*}
\Expect_{\vecw \in (\bbW, \zeta_t)} (S_t(\vecw))
&=& \int_{\bbW}  S_t(\vecw) d\zeta_t(\vecw) \\
&=& \int_{\bbW}  S_t(\vecw) \frac{\Ret_t( S(\vecw))}{\int_{\bbW}\Ret_t( S(\vecw))d\mu(\vecw)}d\mu(\vecw)
= \U_t(S),
\end{eqnarray*}
where the final equality follows from (\ref{equation-universal}).
\end{proof}

In Section~\ref{subsection-sampling}, we show that for certain strategies we can efficiently sample from a distribution $\bar\zeta_t$ that is ``close'' to $\zeta_t$, \ie{} given $\gamma_t > 0$, we generate samples from $\bar\zeta_t$ in $\Oh(\log\frac{1}{\gamma_t})$ time and such that 
\begin{equation}
\label{equation-gamma-t}
\int_{\bbW}\abs{\zeta_t(\vecw) - \bar\zeta_t(\vecw)}d\mu(\vecw) \le \gamma_t.
\end{equation}
Assume for now that we can sample from $\bar\zeta_t$, with $\gamma_t = \frac{\varepsilon^2}{4m(t+1)^4}$, where $\varepsilon$ is the constant appearing in Remark~\ref{remark-S-bdd-below}. Let $\bar\U_t(S) = \int_{\bbW} S_t(\vecw)d\bar\zeta_t(\vecw)$ be the corresponding approximation to $\U(S)$. Lemma~\ref{lemma-close-distribution} tells us that we do not lose much by sampling from $\bar\zeta_t$.

\begin{lemma}
\label{lemma-close-distribution}
For all $n \ge 0$, (1) $\Ret_n(\bar\U(S)) \ge (1 - \varepsilon)\Ret_n(\U(S))$ and (2) if $\U(S)$ is a universalization of $S$, then $\bar\U(S)$ is a universalization of $S$ as well.
\end{lemma}
\begin{proof}
Statement (2) follows directly from (1). To see (1), we need only show that the fraction of wealth we put in each stock $i$ on day $t$ under $\bar\U(S)$ is within a $1 - \frac{\varepsilon}{2(t+1)^2}$ factor of the corresponding amount under $\U(S)$, \ie{} $\bar\U_{ti}(S) \ge (1 - \frac{\varepsilon}{2(t+1)^2})\U_{ti}(S)$ for $0\le t < n$ and $1\le i\le m$. For $\vecw\in\bbW$, let $\gamma_t(\vecw) = |\bar\zeta_t(\vecw) - \zeta_t(\vecw)|$, so that $\int_{\bbW} \gamma_t(\vecw) d\vecw = \gamma_t \le \frac{\varepsilon^2}{4m(t+1)^4}$. We have
\begin{eqnarray*}
\bar\U_{ti}(S) &=& \int_{\bbW}  S_{ti}(\vecw)\bar\zeta_t(\vecw)d\mu(\vecw)
\ge \int_{\bbW}  S_{ti}(\vecw)(\zeta_t(\vecw) - \gamma_t(\vecw))d\mu(\vecw) \\
&=& \U_{ti}(S) - \int_{\bbW} S_{ti}(\vecw)\gamma_t(\vecw) d\mu(\vecw) 
\ge \U_{ti}(S) - \gamma_t \qquad \text{(since $S_{ti}(\vecw) \le 1$)} \\
&\ge& (1 - \textstyle\frac{\varepsilon}{2(t+1)^2})\U_{ti}(S) \quad \text{(since $\U_{ti}(S) \ge \textstyle\min_\vecw S(\vecw) \ge \textstyle\frac{\varepsilon}{2m(t+1)^2}$ and $\gamma_t \le \textstyle\frac{\varepsilon^2}{4m(t+1)^4}$),}
\end{eqnarray*}
as we wanted to show.
\end{proof}

By sampling from $\bar\zeta_t$, we use a generalization of the Chernoff bound to get an approximation $\tilde\U(S)$ to $\bar\U(S)$ such that with high probability $\tilde\U_{ti}(S) \ge (1 - \frac{\varepsilon}{2(t+1)^2})\bar\U_{ti}(S)$ for $0\le t < n$ and $1\le i\le m$. Using an argument similar to that in the proof of Lemma~\ref{lemma-close-distribution}, we see that if $\bar\U(S)$ is a universalization of $S$, then such a $\tilde\U(S)$ is a universalization of $S$ as well. Choose $\vecw_1,\ldots, \vecw_{N_t} \in \bbW$ at random according to distribution $\bar\zeta_t$ and let $\tilde\U_{ti}(S) = \frac{1}{N_t}\sum_{i=1}^{N_t} S_{ti}(\vecw_i)$. Lemma~\ref{lemma-chernoff} discusses the number of samples $N_t$ required to get a sufficiently good approximation to $\bar\U_t(S)$.

\begin{lemma}
\label{lemma-chernoff}
Given $0<\delta<1$ use $N_t \ge \frac{8m^2(t+1)^8}{\varepsilon^4} \log\frac{2m(t+1)^2}{\delta}$ samples to compute $\tilde\U_t(S)$, where $\varepsilon$ is the constant appearing in Remark~\ref{remark-S-bdd-below}. With probability $1-\delta$, $\tilde \U_{ti}(S) \ge (1 - \frac{\varepsilon}{2(t+1)^2})\bar \U_{ti}(S)$ for all $1\le i\le m$ and $t\ge 0$.
\end{lemma}
\begin{proof}
Hoeffding \cite{Hoeffding:1963:PIS} proves a general version of the Chernoff bound. For random variables $0\le X_i \le 1$ with $\Expect(X_i) = \mu$ and $\tilde X = \frac{1}{N}\sum_{i=1}^N X_i$ the bound states that $\Pr(\tilde X \le (1 - \alpha)\mu) \le e^{-2N\alpha^2\mu^2}$. In our case, we would like $\tilde\U_{ti} \ge (1 - \frac{\varepsilon}{2(t+1)^2})\bar\U_{ti}$. As this must hold for $1\le i\le m$ and $t \ge 0$ with total probability $1 - \delta$, we require $\Pr(\tilde\U_{ti} \le (1 - \frac{\varepsilon}{2(t+1)^2})\bar\U_{ti}) \le \frac{\delta}{2m(t+1)^2}$ for each $i$ and $t$. From our assumption stated in Remark~\ref{remark-S-bdd-below}, $\mu = \bar\U_{ti} \ge \frac{\varepsilon}{2m(t+1)^2}$ and the desired probability bound is achieved with $N_t \ge \frac{8m^2(t+1)^8}{\varepsilon^4} \log\frac{2m(t+1)^2}{\delta}$ samples.
\end{proof}

\subsection{Efficient Sampling}
\label{subsection-sampling}

We now discuss how to sample from $\bbW = \calW_k^\ell = \calW_k \times \cdots \times \calW_k$ according to distribution $\zeta_t(\cdot) \propto \Ret_t(\cdot) = \Ret_t( S(\cdot))$. $\bbW$ is a convex set of diameter $d = \sqrt{2\ell}$. We focus on a discretization of the sampling problem. Choose an orthogonal coordinate system on each $\calW_k$ and partition it into hypercubes of side length $\delta_t$, where $\delta_t$ is a constant chosen below. Let $\Omega$ be the set of centers of cubes that intersect $\bbW$ and choose the partition such that the coordinates of $\vecw\in\Omega$ are multiples of $\delta_t$. For $\vecw\in\Omega$, let $C(\vecw)$ be the cube with center $\vecw$. We show how to choose $\vecw\in\Omega$ with probability ``close to'' 
$$
\pi_t(\vecw) = \frac{\Ret_t(\vecw)}{\sum_{\vecw\in\Omega}\Ret_t(\vecw)}.
$$ 
In particular, we sample from a distribution $\tilde\pi_t$ that satisfies
\begin{equation}
\label{equation-discrete-gamma}
\sum_{\vecw\in\Omega} \abs{\pi_t(\vecw) - \tilde\pi_t(\vecw)} \le \gamma_t = \frac{\varepsilon^2}{4m(t+1)^4}.
\end{equation}
Note that this is a discretization of (\ref{equation-gamma-t}). We will also have that for each $\vecw\in\Omega$, 
\begin{equation}
\label{equation-small-relative-error}
\frac{\tilde\pi_t(\vecw)}{\pi_t(\vecw)} \le 2.
\end{equation}
We would like to choose $\delta_t$ sufficiently small that $\Ret_t$ is ``nearly constant'' over $C(\vecw)$ \ie{} there is a small constant $\nu > 0$ such that
\begin{equation}
\label{equation-lipschitz}
(1+\nu)^{-1}\Ret_t(\vecw) \le \Ret_t(\vecw') \le (1+\nu) \Ret_t(\vecw)
\end{equation}
for all $\vecw' \in C(\vecw)$. Such a $\delta_t$ can be chosen for investment strategies $S$ that have bounded derivative, as we see in Lemma~\ref{lemma-choose-delta}.

\begin{lemma}
\label{lemma-choose-delta}
Suppose that investment strategy $S$ satisfies the condition for universalizability given in Theorem~\ref{theorem-main-universalization}, \ie{} $\abs{\frac{\partial S_{ti}(\vecw)}{\partial w_j}} \le ct$. Given $\nu > 0$, let $\delta_t = \delta_t(\nu) = \frac{\nu}{3c'mt^4k\ell}$, where $c'$ is defined in the proof of Theorem~\ref{theorem-main-universalization}. For $\vecw,\vecw'\in\bbW$ such that $|w_{ij} - w'_{ij}| \le \delta_t(\nu)$ for all $1\le i\le \ell$ and $1\le j\le k$, $(1+\nu)^{-1}\Ret_t(\vecw) \le \Ret_t(\vecw') \le (1+\nu) \Ret_t(\vecw)$.
\end{lemma}
\begin{proof}
Note that $|\vecw - \vecw'| \le \delta_t\sqrt{k\ell}$. Let $\vecw^*$ be the parameters that maximize the return on the line between $\vecw$ and $\vecw'$. By the multivariate mean value theorem and the bound for $|\nabla\Ret_t|$ given in (\ref{equation-gradient}),
\begin{eqnarray*}
\Ret_t(\vecw^*) &=& \Ret_t(\vecw) + \Ret_t(\vecw^*) - \Ret_t(\vecw) \\
&\le& \Ret_t(\vecw) + |\nabla\Ret_t(\vecw_m)| \cdot |\vecw - \vecw^*| 
\quad \text{(for some $\vecw_m$ between $\vecw^*$ and $\vecw$)}\\
&\le& \Ret_t(\vecw) + c'\Ret_t(\vecw_m) mn^4\sqrt{k\ell} \cdot \delta_t\sqrt{k\ell} \le \Ret_t(\vecw) + \Ret_t(\vecw^*)\frac{\nu}{3} \\
\Rightarrow\quad \Ret_t(\vecw) &\ge& \Ret_t(\vecw^*)(1 - \frac{\nu}{3})
\ge \Ret_t(\vecw')(1 - \frac{\nu}{3})
\end{eqnarray*}
so that $\Ret_t(\vecw') \le (1 + \nu)\Ret_t(\vecw)$. By similar reasoning,
\begin{eqnarray*}
\Ret_t(\vecw') &=& \Ret_t(\vecw^*) +  \Ret_t(\vecw') - \Ret_t(\vecw^*) \\
&\ge& \Ret_t(\vecw^*) - |\nabla\Ret_t(\vecw_m)| \cdot |\vecw' - \vecw^*| 
  \quad\text{(for some $\vecw_m$ between $\vecw^*$ and $\vecw'$)} \\
&\ge& \Ret_t(\vecw^*)(1 - \frac{\nu}{3})
\ge \Ret_t(\vecw)(1 - \frac{\nu}{3}) \ge \Ret_t(\vecw)(1 + \nu)^{-1},
\end{eqnarray*}
completing the proof.
\end{proof}

We use a Metropolis algorithm \cite{Metropolis:1953:ESC} to sample from $\tilde\pi_t$. We generate a random walk on $\Omega$ according to a Markov chain whose stationary distribution is $\pi_t$. Begin by selecting a point $\vecw_0 \in \Omega$ according to either $\tilde\pi_{t-1}$ or $\tilde\pi_{t-2}$;\footnote{Ideally, we would like to begin with a point selected according to $\tilde\pi_{t-1}$, but, as discussed in Remark~\ref{remark-select-previous}, this is not always possible.} Remark~\ref{remark-select-previous} explains how to do this.

\begin{remark}
\label{remark-select-previous}
We can select a point according to $\tilde\pi_{t-1}$ by ``saving'' our samples that were generated at time $t-1$. By Lemma~\ref{lemma-chernoff}, we would have generated $N_{t-1} \ge \frac{8m^2t^8}{\varepsilon^4} \log\frac{2mt^2}{\delta}$ samples at time $t-1$, which is not enough to generate the $N_t \ge \frac{8m^2(t+1)^8}{\varepsilon^4} \log\frac{2m(t+1)^2}{\delta}$ samples necessary at time $t$. Instead, we can ``save'' samples that were generated at times $t-1$ and $t-2$. For sufficiently large $t$, $N_t \le N_{t-1} + N_{t-2}$ and our initial point $\vecw_0$ would be picked according to either $\tilde\pi_{t-1}$ or $\tilde\pi_{t-2}$. As we see in the proof of Lemma~\ref{lemma-M}, this distinction is not important.
\end{remark}

If $\vecw_\tau$ is the position of our random walk at time $\tau \ge 0$, we pick its position at time $\tau + 1$ as follows. Note that $\vecw_\tau$ has $2(k-1)\ell$ neighbors, two along each axis in the Cartesian product of $\ell$ $(k-1)$-dimensional spaces. Let $\vecw$ be a neighbor of $\vecw_\tau$, selected uniformly at random. If $\vecw\in\Omega$, set
$$
\vecw_{\tau+1} = 
\begin{cases}
\vecw & \text{with probability $p = \min(1, \frac{\Ret_t(\vecw)}{\Ret_t(\vecw_\tau)})$} \\
\vecw_\tau & \text{with probability $1 - p$.}
\end{cases}
$$
If $\vecw\not\in\Omega$, let $\vecw_{\tau+1} = \vecw_{\tau}$. It is well-known that the stationary distribution of this random walk is $\pi_t$. We must determine how many steps of the walk are necessary before the distribution has gotten sufficiently close to stationary. Let $p_{\tau}$ be the distribution attained after $\tau$ steps of the random walk. That is, $p_\tau(\vecw)$ is the probability of being at $\vecw$ after $\tau$ steps. 

\begin{remark}
A distinction should be made between $t$ and $\tau$. We use $t$ to refer to the time step in our universalization algorithm. We use $\tau$ to refer to ``sub'' time steps used in the Markov chain to sample from $\pi_t$. When $t$ is clear from context, we may drop it from the subscripts in our notation.
\end{remark}

Applegate and Kannan \cite{Applegate:1991:SIN} show that if the desired distribution $\pi_t$ is proportional to a log-concave function $F$ (\ie{} $\log F$ is concave) the Markov chain is \emph{rapidly mixing}, reaches its steady state in polynomial time. Frieze and Kannan \cite{Frieze:1999:LSI} give an improved upper bound on the mixing time using Logarithmic Sobolev inequalities \cite{Diaconis:1996:LSI}.

\begin{theorem}[Theorem 1 of \cite{Frieze:1999:LSI}]
\label{theorem-kannan}
Assume the diameter $d$ of $\bbW$ satisfies $d\ge \delta_t \sqrt{k\ell}$ and that the target distribution $\pi$ is proportional to a log-concave function. There is an absolute constant $\kappa > 0$ such that
\begin{equation}
\label{equation-tv-distance}
2\left(\sum_{\vecw\in\Omega}|\pi(\vecw) - p_\tau(\vecw)|\right)^2 \le 
e^{-\frac{\kappa\tau\delta_t^2}{k\ell d^2}}\log\frac{1}{\pi_*} +
\frac{M\pi_e k\ell d^2}{\kappa\delta_t^2},
\end{equation}
where $\pi_* = \min_{\vecw\in\Omega}\pi(\vecw)$,
$M = \max_{\vecw\in\Omega}\frac{p_0(\vecw)}{\pi(\vecw)}\log\frac{p_0(\vecw)}{\pi(\vecw)}$, $p_0(\cdot)$ is the initial distribution on $\Omega$,
$\pi_e = \sum_{\vecw \in \Omega_e} \pi(\vecw)$, and
$\Omega_e = \Set{\vecw\in\Omega}{\Vol(C(\vecw) \cap \bbW) < \Vol(C(\vecw))}$ (the ``$e$'' in the subscripts of $\pi_e$ and $\Omega_e$ stands for ``edge'').
\end{theorem}

In the random walk described above, if $\vecw_\tau$ is on an edge of $\Omega$, so it has many neighbors outside $\Omega$, the walk may get ``stuck'' at $\vecw_\tau$ for a long time, as seen in the ``$\pi_e$'' term of Theorem~\ref{theorem-kannan}. We must ensure that the random walk has low probability of reaching such edge points. We do this by applying a ``damping function'' to $\Ret_t$ that becomes exponentially small near the edges of $\bbW$. For $1\le i\le \ell$, $1\le j\le k$, and $\vecw = (\vecw_1, \ldots, \vecw_\ell) = ((w_{11}, \ldots, w_{1k}), \ldots, (w_{\ell 1}, \ldots, w_{\ell k})) \in\bbW$ let
\begin{equation}
\label{equation-fij}
f_{ij}(\vecw) = e^{\Gamma\min(-\sigma + w_{ij}, 0)},
\end{equation}
where $\sigma > 0$ and $\Gamma > 2$ are constants that we choose below, and let
$$
F_t(\vecw) = \Ret_t(\vecw)\prod_{i=1}^\ell\prod_{j=1}^k f_{ij}(\vecw).
$$

\begin{lemma}
$F_t$ is log-concave if and only if $\Ret_t$ is log-concave.\footnote{We characterize investment strategies for which $\Ret_t$ is log-concave in Theorem~\ref{theorem-log-concave}}
\end{lemma}
\begin{proof}
This follows from the fact that log-concave functions are closed under multiplication and the fact that $\log f_{ij}(\vecw) = \Gamma\min(-\sigma + w_{ij}, 0)$, which is concave.
\end{proof}

Choose $\sigma = \frac{1}{k}\delta_t(\frac{\gamma_t}{2})$, where $\delta_t(\cdot)$ is defined in Lemma~\ref{lemma-choose-delta} and $\gamma_t$ is defined in (\ref{equation-discrete-gamma}). Let $\zeta_F\propto F_t$ be the probability measure proportional to $F_t$. We need to show that for our purposes, sampling from $\zeta_F$ is not much different than sampling from $\zeta_t$. By Lemma~\ref{lemma-close-distribution}, we can do this by showing that $\int_\bbW |\zeta_t(\vecw) - \zeta_F(\vecw)|d\vecw \le \gamma_t$, which we do in Lemma~\ref{lemma-F-not-too-different}. 

\begin{remark}
\label{remark-scale-W}
Before continuing, we show how $\bbW$ can be scaled, which will be useful in future proofs. Take $\vecp = (\frac{1}{k}, \ldots, \frac{1}{k}) \in \calW_k$; given $\chi \in (-1, 1)$, let
$$
\vecw^{(\chi)} = (1 + \chi)(\vecw - \vecp) + \vecp
$$
and let
$$
\calW_k^{(\chi)} = \Set{\vecw^{(\chi)}}{\vecw\in\calW_k}
$$
be a scaled version of $\calW_k$ about $\vecp$, where the scaling factor is $1+\chi$. To extend this scaling to $\bbW = \calW_k^\ell$, given $\vecw = (\vecw_1, \ldots, \vecw_\ell) \in \bbW$, let $\vecw^{(\chi)} = (\vecw_1^{(\chi)}, \ldots, \vecw_\ell^{(\chi)})$ and let 
$$
\bbW^{(\chi)} = \Set{\vecw^{(\chi)}}{\vecw\in\bbW}.
$$
A fact we use is that for $1\le i\le \ell$, $1\le j\le k$, and $\vecw = (\vecw_1, \ldots, \vecw_\ell) \in \bbW$
$$
|w_{ij}^{(\chi)} - w_{ij}| 
= |(1+\chi)(w_{ij} - \frac{1}{k}) + \frac{1}{k} - w_{ij}|
\le |\chi|.
$$
\end{remark}

\begin{lemma}
\label{lemma-F-not-too-different}
$\int_\bbW |\zeta_t(\vecw) - \zeta_F(\vecw)|d\vecw \le \gamma_t$.
\end{lemma}
\begin{proof}
Let $\bbW' = \bbW^{(-k\sigma)}$ be the ``scaled-in'' version of $\bbW$, as defined in Remark~\ref{remark-scale-W}. By Lemma~\ref{lemma-choose-delta}, since $|w_{ij} - w_{ij}'| \le k\sigma = \delta_t(\frac{\gamma_t}{2})$ for all $i$ and $j$, $\Ret_t(\vecw') \ge \frac{1}{1 + \frac{\gamma_t}{2}} \Ret_t(\vecw)$ and 
\begin{equation}
\label{equation-Wprime-close}
\int_{\bbW'}\Ret_t(\vecw)d\vecw \ge \frac{1}{1 + \frac{\gamma_t}{2}} \int_{\bbW}\Ret_t(\vecw)d\vecw.
\end{equation}
Let $\bbW_{eq} = \Set{\vecw\in\bbW}{F_t(\vecw) = \Ret_t(\vecw)}$ be the subset of $\bbW$ where $F_t(\cdot)$ and $\Ret_t(\cdot)$ are equal; $\bbW' \subset \bbW_{eq}$ since, by construction of $\vecw'$, $w'_{ij} \ge \sigma$ for all $i$ and $j$. Let 
$\bbW_{+} = \Set{\vecw\in\bbW}{\zeta_F(\vecw) \ge \zeta_t(\vecw)}$ be the subset of $\bbW$ where $\zeta_F(\cdot)$ is at least $\zeta_t(\cdot)$ and let $\bbW_{-} = \bbW - \bbW_{+}$. We bound
$$
\int_\bbW |\zeta_F(\vecw) - \zeta_t(\vecw)|d\vecw = \int_{\bbW_+}(\zeta_F(\vecw) - \zeta_t(\vecw))d\vecw + \int_{\bbW_-}(\zeta_t(\vecw) - \zeta_F(\vecw))d\vecw
$$
by bounding $\int_{\bbW_-}(\zeta_t - \zeta_F)$, which also gives a bound for $\int_{\bbW_+}(\zeta_F - \zeta_t)$, since 
$$
\int_{\bbW_+}(\zeta_F - \zeta_t) = \left(1 - \int_{\bbW_-}\zeta_F\right) - \left(1 - \int_{\bbW_-}\zeta_t\right) = \int_{\bbW_-}(\zeta_t - \zeta_F).
$$ 
Since $F_t \le \Ret_t$, $\int_\bbW F_t \le \int_\bbW \Ret_t$ and 
$\zeta_F(\vecw) = \frac{F_t(\vecw)}{\int_\bbW F_t} \ge \frac{\Ret_t(\vecw)}{\int_{\bbW}\Ret_t} = \zeta_t(\vecw)$ for $\vecw\in\bbW_{eq}$; thus $\bbW'\subset\bbW_{eq}\subset \bbW_+$ and $\bbW_-\subset \bbW - \bbW'$.
We have
\begin{eqnarray*}
\int_{\bbW_-}(\zeta_t(\vecw) - \zeta_F(\vecw))d\vecw
&\le& \int_{\bbW - \bbW'}\zeta_t(\vecw)d\vecw
= \frac{\int_{\bbW - \bbW'}\Ret_t(\vecw)d\vecw}{\int_{\bbW}\Ret_t(\vecw)d\vecw}
= 1 - \frac{\int_{\bbW'}\Ret_t(\vecw)d\vecw}{\int_{\bbW}\Ret_t(\vecw)d\vecw} \\
&\le& 1 - \frac{1}{1+ \frac{\gamma_t}{2}} \le \frac{\gamma_t}{2},
\end{eqnarray*}
where the second-last inequality follows from (\ref{equation-Wprime-close}). This completes the proof. 
\end{proof}

Henceforth, we are concerned with sampling from $\bbW$ with probability proportional to $F_t(\cdot)$. We use the Metropolis algorithm described above, replacing $R_t(\cdot)$ with $F_t(\cdot)$; we must refine our grid spacing $\delta_t$ so that (\ref{equation-lipschitz}) is satisfied by $F_t$; let $\delta_t'$ be the new grid spacing.

\begin{lemma}
\label{lemma-lipschitz}
Suppose that the conditions of Lemma~\ref{lemma-choose-delta} are satisfied. Given $\nu > 0$, let $\delta_t'(\nu) = \delta_t' = \frac{\nu}{3\Gamma c'mt^4k\ell} = \delta_t(\frac{\nu}{\Gamma})$, where $\Gamma$ appears in (\ref{equation-fij}). For $\vecw,\vecw'\in\bbW$ such that $|w_{ij} - w'_{ij}| \le \delta_t'(\nu)$ for all $1\le i\le \ell$ and $1\le j\le k$, $(1+\nu)^{-1} F_t(\vecw) \le F_t(\vecw') \le (1+\nu) F_t(\vecw)$.
\end{lemma}
\begin{proof}
By Lemma~\ref{lemma-choose-delta}, $\Ret_t(\vecw)$ and $\Ret_t(\vecw')$ differ by at most a factor $1 + \frac{\nu}{\Gamma}$. For each $i$ and $j$, $f_{ij}(\vecw)$ and $f_{ij}(\vecw')$ differ by at most a factor $e^{\Gamma\delta_t'(\nu)}$ and hence $\prod_{i=1}^\ell\prod_{j=1}^k f_{ij}(\vecw)$ and $\prod_{i=1}^\ell\prod_{j=1}^k f_{ij}(\vecw')$ differ by at most a factor $e^{k\ell\Gamma\delta_t'(\nu)} = e^{\frac{\nu}{3c'mt^4}}$. Hence, for $\Gamma \ge 2$ and sufficiently large $t$, $F_t(\vecw)$ and $F_t(\vecw')$ differ by at most a factor $1 + \nu$.
\end{proof}

We are now ready to use Theorem~\ref{theorem-kannan} to select $\tau$ so that the resulting distribution $p_\tau$ satisfies (\ref{equation-discrete-gamma}) (Theorem~\ref{theorem-kannan-corollary}) and (\ref{equation-small-relative-error}) (Theorem~\ref{theorem-small-relative-error}), with $p_\tau$ in place of $\tilde\pi_t$ and $F_t$ in place of $\Ret_t$. We begin with some preliminary lemmas.

\begin{lemma}
\label{lemma-pi-star}
There is a constant $\beta > 0$ such that $\log\frac{1}{\pi_*} \le k\ell\Gamma\sigma + k\ell\log\frac{\beta}{\delta_t'} + t\log\frac{2mt^2}{\varepsilon}$, where $\varepsilon$ is defined in Remark~\ref{remark-S-bdd-below}.
\end{lemma}
\begin{proof}
Take $\beta$ such that the number of points in $\Omega$ is at most $(\frac{\beta}{\delta_t'})^{(k-1)\cdot \ell}$.  For $\vecw_1, \vecw_2 \in\Omega$, the ratio of single-day returns on day $t'$ using $\vecw_1$ and $\vecw_2$ is
$$
\frac{S_{t'}(\vecw_1)\cdot \vecx_{t'}}{S_{t'}(\vecw_2) \cdot \vecx_{t'}} 
\ge \frac{\varepsilon}{2m(t'+1)^2}
$$
by Remark~\ref{remark-S-bdd-below} and Lemma~\ref{lemma-vector-ratio}. The ratio of the cumulative returns up to day $t$ is
$$
\frac{\Ret_t(\vecw_1)}{\Ret_t(\vecw_2)} 
\ge \left( \frac{\varepsilon}{2mt^2} \right)^t,
$$
and thus $\frac{\Ret_t(\vecw)}{\sum_{\vecw\in\Omega}\Ret_t(\vecw)} \ge (\frac{\delta_t'}{\beta})^{(k-1)\ell} \left( \frac{\varepsilon}{2mt^2} \right)^t$. Factoring in the maximum dampening effect of the $f_{ij}$, 
$\pi_* \ge e^{-k\ell\Gamma\sigma} (\frac{\delta_t'}{\beta})^{(k-1)\ell} \left( \frac{\varepsilon}{2mt^2} \right)^t$ 
and 
$\log\frac{1}{\pi_*} \le k\ell\Gamma\sigma + k\ell\log\frac{\beta}{\delta_t'} + t\log\frac{2mt^2}{\varepsilon}$.
\end{proof}

\begin{lemma}
\label{lemma-M}
$M \le 4\left(\frac{2m(t+1)^2}{\varepsilon}\right)^2\log\frac{2m(t+1)^2}{\varepsilon}$.
\end{lemma}
\begin{proof}
As stated in Remark~\ref{remark-select-previous}, the initial distribution is either $p_0 = \tilde\pi_{t-1}$ or $\tilde\pi_{t-2}$. It turns out that the worst case happens when $p_0 = \tilde\pi_{t-2}$. For all $\vecw\in\Omega$, $\frac{\tilde\pi_{t-2}(\vecw)}{\pi_{t-2}(\vecw)} \le 2$ by (\ref{equation-small-relative-error}) and 
\begin{eqnarray*}
\frac{\pi_{t-2}(\vecw)}{\pi_t(\vecw)} 
&=& \frac{F_{t-2}(\vecw)}{\sum_{\vecw\in\Omega}F_{t-2}(\vecw)} 
   \cdot \frac{\sum_{\vecw\in\Omega}F_{t}(\vecw)}{F_{t}(\vecw)} \\
&\le& \frac{F_{t-2}(\vecw)}{F_{t}(\vecw)} 
   \cdot \frac{F_{t}(\vecw')}{F_{t-2}(\vecw')} \quad\text{(by Lemma~\ref{lemma-vector-ratio}, where $\vecw' = \arg\max_{\vecw\in\Omega}\frac{F_{t}(\vecw)}{F_{t-2}(\vecw)}$)} \\
&=& \frac{\Ret_{t-2}(\vecw)}
         {\Ret_{t}(\vecw)} \cdot 
    \frac{\Ret_{t}(\vecw')}
         {\Ret_{t-2}(\vecw')} \quad\text{(since the $\{f_{ij}(\cdot)\}_{i,j}$ remain constant with time)} \\
&=& \frac{(S_t(\vecw')\cdot\vecx_t)(S_{t-1}(\vecw')\cdot\vecx_{t-1})}
       {(S_t(\vecw)\cdot\vecx_t)(S_{t-1}(\vecw)\cdot\vecx_{t-1})}
\le \left(\frac{2m(t+1)^2}{\varepsilon}\right)^2,
\end{eqnarray*}
where the final inequality follows from the discussion in the proof of Lemma~\ref{lemma-pi-star}. This proves the result since $\frac{\tilde\pi_{t-2}(\vecw)}{\pi_t(\vecw)} = \frac{\tilde\pi_{t-2}(\vecw)}{\pi_{t-2(\vecw)}} \frac{\pi_{t-2(\vecw)}}{\pi_t(\vecw)}$.
\end{proof}

\begin{lemma}
\label{lemma-pi-theta}
$\pi_e \le (1 + \nu)^4(1 + \frac{\gamma_t}{2})e^{-\Gamma\sigma}$, where 
$\nu$ appears in the definition of $\delta_t'$ in Lemma~\ref{lemma-lipschitz},
$\gamma_t$ appears in (\ref{equation-discrete-gamma}), and
$\Gamma$ and $\sigma$ appear in (\ref{equation-fij}).
\end{lemma}
\begin{proof}
Extend our $\delta_t'$-hypercube partition of $\bbW$ to the hyperplane containing $\bbW$ and let $\Psi$ be the set of centers of the hypercubes in this extended partition. For $K \subset \R^{k\ell}$, let $\Psi_K$ be the set of grid points $\vecw\in\Psi$ such that $C(\vecw) \cap K \not=\emptyset$, so that $\Omega = \Psi_\bbW$. By Lemma~\ref{lemma-lipschitz}, for $K\subset \bbW$,
\begin{equation}
\label{equation-sum-to-int}
\frac{1}{1 + \nu}\sum_{\vecw\in\Psi_K}F_t(\vecw)\Vol(C(\vecw)\cap K) 
\le \int_K F_t(\vecw)d\vecw
\le (1 + \nu)\sum_{\vecw\in\Psi_K}F_t(\vecw)\Vol(C(\vecw)\cap K).
\end{equation}
Using the notation of Lemma~\ref{lemma-F-not-too-different}, let $\bbW' = \bbW^{(-k\sigma)}$ be a ``scaled-in'' version of $\bbW$; we showed in Lemma~\ref{lemma-F-not-too-different} that for $\vecw\in\bbW'$, $F_t(\vecw) = \Ret_t(\vecw)$ and that
\begin{equation}
\label{equation-W'}
\int_{\bbW'}F_t(\vecw)d\vecw = \int_{\bbW'}\Ret_t(\vecw)d\vecw \ge \frac{1}{1 + \frac{\gamma_t}{2}} \int_{\bbW}\Ret_t(\vecw)d\vecw.
\end{equation}
Let $\bbW'' = \bbW^{(\delta_t'(\nu))}$ be a ``scaled-out'' version of $\bbW$ and extend the domains of $F_t(\cdot)$ and $\Ret_t(\cdot)$ to $\bbW''$ by defining $F_t(\vecw'') = F_t(\bar\vecw'')$ and $\Ret_t(\vecw'') = \Ret_t(\bar\vecw'')$ for $\vecw'' \in\bbW'' - \bbW$, where $\bar\vecw''$ is the point where the line between $\vecw''$ and $\vecp^\ell = (\vecp, \ldots, \vecp) \in \bbW$ intersects the boundary of $\bbW$. By Lemma~\ref{lemma-lipschitz} and the construction of the extension of $\Ret_t$, $\Ret_t(\vecw'') \le (1 + \nu)\Ret_t(\vecw)$ and
\begin{equation}
\label{equation-W''}
\int_{\bbW''}\Ret_t(\vecw)d\vecw \le (1 + \nu)\int_{\bbW}\Ret_t(\vecw)d\vecw.
\end{equation}
By construction of $\bbW''$, $C(\vecw) \subset \bbW''$ for $\vecw\in\Omega_e$; from the definition of $F_t$ and the choice of $\delta_t'$, $F_t(\vecw) \le (1+\nu)e^{-\Gamma\sigma}\Ret_t(\vecw)$ for $\vecw\in\Omega_e$. Using these facts,
\begin{eqnarray*}
\pi_e &=& \frac{\sum_{\vecw\in\Omega_e} F_t(\vecw)}{\sum_{\vecw\in\Omega} F_t(\vecw)}
\le \frac{\delta_t^{(k-1)\ell}}{\delta_t^{(k-1)\ell}} \cdot\frac{(1 + \nu)e^{-\Gamma\sigma}\sum_{\vecw\in\Omega_e} \Ret_t(\vecw)}{\sum_{\vecw\in\Omega} F_t(\vecw)} \\
&\le& (1 + \nu)e^{-\Gamma\sigma} \frac{\sum_{\vecw\in\Psi_{\bbW''}}\Vol(C(\vecw)\cap\bbW'') \Ret_t(\vecw)}
   {\sum_{\vecw\in\Psi_{\bbW}}\Vol(C(\vecw)\cap\bbW) F_t(\vecw)} \quad
    \text{(since $\Vol(C(\vecw)) = \delta_t^{(k-1)\ell}$)} \\
&\le& (1 + \nu)e^{-\Gamma\sigma} 
  \frac{(1 + \nu)\int_{\bbW''}\Ret_t(\vecw)d\vecw}
       {\frac{1}{(1+\nu)}\int_{\bbW}F_t(\vecw)d\vecw}
\quad\text{(by (\ref{equation-sum-to-int}))} \\
&\le& (1 + \nu)^3e^{-\Gamma\sigma} 
  \frac{\int_{\bbW''}\Ret_t(\vecw)d\vecw}{\int_{\bbW'}F_t(\vecw)d\vecw}
\le (1 + \nu)^4(1 + \frac{\gamma_t}{2})e^{-\Gamma\sigma} 
    \quad\text{(by (\ref{equation-W'}) and (\ref{equation-W''})).}
\end{eqnarray*}
\end{proof}

\begin{remark}
\label{remark-simple-notation}
We simplify notation below by using $\Oh^*(\cdot)$ notation, which ignores logarithmic and constant terms. For our purposes, $f(\cdot) = \Oh^*(g(\cdot))$ if there exists a constant $C \ge 0$ such that $f(\cdot) = \Oh(g(\cdot) \log^C(k\ell m t/\varepsilon))$. The values derived above in this notation are 
$\gamma_t = \Oh^*(\frac{\varepsilon^2}{mt^4})$, 
$\delta_t = \Oh^*(\frac{\nu}{mt^4k\ell})$, 
$\sigma = \Oh^*(\frac{\varepsilon^2}{m^2t^8k^2\ell})$,
$\delta_t' = \Oh^*(\frac{\nu}{\Gamma mt^4k\ell})$,
$\log\frac{1}{\pi_*} = \Oh^*(k\ell\Gamma\sigma + t)$,
$M = \Oh^*(\frac{m^2t^4}{\varepsilon^2})$, and
$\pi_e = \Oh^*(e^{-\Gamma\sigma})$.
\end{remark}

\begin{theorem}
\label{theorem-kannan-corollary}
Letting $\Gamma = \Oh^*(\frac{1}{\sigma}) = \Oh^*(\frac{m^2t^8k^2\ell}{\varepsilon^2})$,
the random walk reaches a distribution $\tilde \pi$ that satisfies (\ref{equation-discrete-gamma}) after $\tau = \Oh^*(\frac{k^7\ell^6 m^6 t^{24}}{\kappa \nu^2 \varepsilon^4})$ steps.
\end{theorem}
\begin{proof}
We show how to bound the right-side of (\ref{equation-tv-distance}), where the grid spacing $\delta_t$ has been replaced by $\delta_t'$. The second term, $\frac{M\pi_e k\ell d^2}{\kappa{\delta_t'}^2}$, can be made exponentially small in $\Gamma$ by choosing $\Gamma = \Oh^*(\frac{1}{\sigma})$. The value of $\tau$ stated in the theorem is large enough to make the first term, $e^{-\frac{\kappa\tau{\delta_t'}^2}{k\ell d^2}}\log\frac{1}{\pi_*}$, exponentially small in $\tau$.
\end{proof}

\begin{theorem}
\label{theorem-small-relative-error}
Suppose that the distribution $p_{\tau_0}$ obtained after $\tau_0$ steps satisfies
$$
\sum_{\vecw\in\Omega} \abs{\pi(\vecw) - p_{\tau_0}(\vecw)} \le \gamma_t.
$$
After $\tau_0' \ge \frac{\tau_0}{\tau_0 - \log\frac{1}{\pi_*} - \log\frac{1}{\gamma_t}}\log\frac{1}{\pi_*} = \Oh^*(\tau_0(k\ell + t))$ steps, the resulting distribution $p_{\tau_0'}$ satisfies 
$$
\max_{\vecw\in\Omega}\frac{p_{\tau_0'}(\vecw)}{\pi(\vecw)} - 1 \le 1,
$$
which implies (\ref{equation-small-relative-error}).
\end{theorem}
\begin{proof}
Let $d(\tau) = \frac{1}{2}\sum_{\vecw\in\Omega} \abs{\pi(\vecw) - p_\tau(\vecw)}$ and $\hat d(\tau) = \max_{\vecw\in\Omega}\frac{p_{\tau}(\vecw)}{\pi(\vecw)} - 1$ so that $d(\tau_0) \le \frac{1}{2}\gamma_t$. Aldous and Fill prove \cite[Equations (5) and (6)]{Aldous:1999:ALT} that if 
$\tau \ge \frac{1}{\lambda}\log\frac{1}{\pi_*}$, then $\hat d(\tau) \le 1$, where $\pi_* = \min_{w\in\Omega}\pi_t(w)$ is as defined in the statement of Theorem~\ref{theorem-kannan} and $\lambda$ is the second-largest eigenvalue of the steady-state transition matrix $P$ of $\pi_t$.

To prove the bound on $\tau_0'$, we show that $\lambda \ge \frac{\tau_0 - \log\frac{1}{\pi_*} - \log\frac{1}{\gamma_t}}{\tau_0} = 1 - \frac{\log\frac{1}{\pi_*} + \log\frac{1}{\gamma_t}}{\tau_0}$. We do this by appealing to a result from Sinclair \cite[Proposition 1 (i)]{Sinclair:1992:IBM}, which states that
$$
\tau_0 \le \frac{\log\frac{1}{\pi_*} + \log\frac{1}{\gamma_t}}{1 - \lambda}.\footnote{Strictly speaking, this result pertains to $\lambda_{\max}$, the second-largest absolute value of the eigenvalues of $P$, but as Sinclair discusses \cite[Page 355]{Sinclair:1992:IBM} the smallest eigenvalue is unimportant, as $P$ can be modified so that all eigenvalues are positive without affecting mixing times beyond a constant factor.}
$$
Solving for $\lambda$ yields the bound for $\tau_0'$. The $\Oh^*(\cdot)$ bound comes from the fact that $\Gamma\sigma = \Oh^*(1)$ and that $\log\frac{1}{\gamma_t}$ and $\log\frac{1}{\pi_*}$ are low-order terms relative to the $\tau_0$ obtained in Theorem~\ref{theorem-kannan-corollary}.
\end{proof}

\subsection{Application to Investment Strategies}

The efficient sampling techniques of this section are applicable to investment strategies $S$ whose return functions $\Ret_n(S(\cdot))$ are log-concave. Theorem~\ref{theorem-log-concave} and Corollary~\ref{corollary-log-concave} characterize such functions.

\begin{theorem}
\label{theorem-log-concave} 
Given investment strategy $S$, suppose that for all parameters $w_i$ and $w_j$, $\frac{\partial^2S}{\partial w_i\partial w_j} = 0$. Then $\Ret_t(\vecw) = \Ret_t(S(\vecw))$ is log-concave.
\end{theorem}
\begin{proof}
Let $r_t(\vecw) = S_t(\vecw)\cdot \vecx_t$, so that $\Ret_n(\vecw) = \prod_{t=0}^{n-1}r_t(\vecw)$. Since log-concave functions are closed under multiplication, we need only show that $r_t(\vecw)$ is log-concave. The gradient vector of $\log r_t(\vecw)$ has $i$-th element 
$\frac{\partial \log r_t(\vecw)}{\partial w_i} 
= \frac{1}{r_t(\vecw)} \frac{\partial r_t(\vecw)}{\partial w_i}$ and the
matrix of second derivatives has $(i,j)$-th element 
$$
-\frac{1}{r_t(\vecw)^2} 
\frac{\partial r_t(\vecw)}{\partial w_i} 
\frac{\partial r_t(\vecw)}{\partial w_j} 
+ \frac{1}{r_t(\vecw)}\frac{\partial^2 r_t(\vecw)}{\partial w_i \partial w_j}
= -\frac{1}{r_t(\vecw)^2} 
\frac{\partial r_t(\vecw)}{\partial w_i} 
\frac{\partial r_t(\vecw)}{\partial w_j}
$$
since 
$\frac{\partial^2 r_t(\vecw)}{\partial w_i \partial w_j} 
= \sum_{\iota = 1}^m \frac{\partial^2 S_{t\iota}(\vecw)}{\partial w_i \partial w_j} \cdot x_{t\iota} = 0$ by assumption. The matrix of second derivatives is negative semidefinite, implying that $\log r_t(\vecw)$ is a concave function.
\end{proof}

\begin{corollary}
\label{corollary-log-concave}
Universalizations of the following investment strategies can be computed using the sampling techniques of this section.
\begin{enumerate}
\item The trading strategies $\MA[k]$ and $\SR[k]$ with long/short allocation functions $g_\ell(x)$ and $h_p(x,y)$ respectively; and 
\item The portfolio strategies $\CRP$ and $\CRPside$.
\end{enumerate}
\end{corollary}
\begin{proof}
The result follows from a straightforward differentiation of the investment descriptions of these strategies.
\end{proof}

\section{Further Research}
\label{section-further-research}

We have introduced in this paper a general framework for universalizing parameterized investment strategies. It would be interesting to see whether the proof of Theorem~\ref{theorem-main-universalization} can be optimized so that existing universal portfolio proofs for $\CRP$ \cite{Cover:1991:UP,Cover:1996:UPS,Blum:1999:UPT} are a special case of Theorem~\ref{theorem-main-universalization}. These proofs not only prove that $\LogRet_n(\U(\CRP))$ converges to $\LogRet_n(\CRP(\vecw^*_n))$, but also prove a bound on the rate of convergence, 
$$
\frac{\Ret_n(\CRP(\vecw^*_n))}{\Ret_n(\U(\CRP))} \le \binom{n+m-1}{m-1} \le (n+1)^{m-1}.
$$
It would also be interesting to study other trading and portfolio strategies that fit in our universalization framework and to see how our universalization algorithms perform in empirical tests.

\bibliographystyle{abbrv}
\bibliography{all}

\begin{thebibliography}{10}

\bibitem{Aldous:1999:ALT}
D.~Aldous and J.~A. Fill.
\newblock Advanced ${L}^2$ techniques for bounding mixing times.
\newblock In {\em Reversible Markov Chains and Random Walks on Graphs}. 1999.
\newblock Unpublished monograph. Available at
  http://stat-www.berkeley.edu/users/aldous/book.html.

\bibitem{Applegate:1991:SIN}
D.~Applegate and R.~Kannan.
\newblock Sampling and integration of near log-concave functions.
\newblock In {\em Proceedings of the 23rd Annual ACM Symposium on Theory of
  Computing}, pages 156--163, 1991.

\bibitem{Blum:1999:UPT}
A.~Blum and A.~Kalai.
\newblock Universal portfolios with and without transaction costs.
\newblock {\em Machine Learning}, 35(3):193--205, 1999.

\bibitem{Brock:1992:STT}
W.~Brock, J.~Lakonishok, and B.~Le{B}aron.
\newblock Simple technical trading rules and the stochastic properties of stock
  returns.
\newblock {\em Journal of Finance}, 47(5):1731--1764, 1992.

\bibitem{Cover:1991:UP}
T.~M. Cover.
\newblock Universal portfolios.
\newblock {\em Mathematical Finance}, 1(1):1--29, Jan. 1991.

\bibitem{Cover:1996:UPS}
T.~M. Cover and E.~Ordentlich.
\newblock Universal portfolios with side information.
\newblock {\em {IEEE} Transactions on Information Theory}, 42(2):348--363,
  March 1996.

\bibitem{Diaconis:1996:LSI}
P.~Diaconis and L.~Saloff-Coste.
\newblock Logarithmic {S}obolev inequalities for finite {M}arkov chains.
\newblock {\em Annals of Applied Probability}, 6:695--750, 1996.

\bibitem{Folland:1984:RAM}
G.~B. Folland.
\newblock {\em Real Analysis: Modern Techniques and their Applications}.
\newblock John Wiley \& Sons, New York, 1984.

\bibitem{Frieze:1999:LSI}
A.~Frieze and R.~Kannan.
\newblock Log-{S}obolev inequalities and sampling from log-concave
  distributions.
\newblock {\em Annals of Applied Probability}, 9:14--26, 1999.

\bibitem{Gartley:1935:PSM}
H.~M. Gartley.
\newblock {\em Profits in the Stock Market}.
\newblock Lambert Gann Publishing Company, Pomeroy, WA, 1935.

\bibitem{Helmbold:1998:LPS}
D.~P. Helmbold, R.~E. Schapire, Y.~Singer, and M.~K. Warmuth.
\newblock On-line portfolio selection using multiplicative updates.
\newblock {\em Mathematical Finance}, 8(4):325--347, 1998.

\bibitem{Hoeffding:1963:PIS}
W.~Hoeffding.
\newblock Probability inequalities for sums of bounded random variables.
\newblock {\em Journal of the American Statistical Association}, 58:13--30,
  March 1963.

\bibitem{Kalai:2000:EAU}
A.~Kalai and S.~Vempala.
\newblock Efficient algorithms for universal portfolios.
\newblock In {\em Proceedings of the 41st Annual IEEE Symposium on Foundations
  of Computer Science}, pages 486--491, 2000.

\bibitem{Metropolis:1953:ESC}
N.~Metropolis, A.~W. Rosenbluth, M.~N. Rosenbluth, A.~H. Teller, and E.~Teller.
\newblock Equation of state calculation by fast computing machines.
\newblock {\em Journal of Chemical Physics}, 21:1087--1092, 1953.

\bibitem{Ordentlich:1996:UPS}
E.~Ordentlich and T.~M. Cover.
\newblock Online portfolio selection.
\newblock In {\em Proceedings of the 9th Annual Conference on Computational
  Learning Theory}, pages 310--313, 1996.

\bibitem{Sinclair:1992:IBM}
A.~Sinclair.
\newblock Improved bounds for mixing rates of {M}arkov chains and
  multicommodity flow.
\newblock {\em Combinatorics, Probability and Computing}, 1:351--370, 1992.

\bibitem{Sullivan:1999:DST}
R.~Sullivan, A.~Timmermann, and H.~White.
\newblock Data-snooping, technical trading rules and the bootstrap.
\newblock {\em Journal of Finance}, 54:1647--1692, 1999.

\bibitem{Wyckoff:1910:STR}
R.~Wyckoff.
\newblock {\em Studies in Tape Reading}.
\newblock Fraser Publishing Company, Burlington, VT, 1910.

\end{thebibliography}

\end{document}